%% file: main.tex
\documentclass[sigconf,nonacm,balance=false]{acmart}

\AtBeginDocument{%
  \providecommand\BibTeX{{%
    \normalfont B\kern0.5em{\scshape i\kern0.25em b}\kern0.8em\TeX}}}

\newif\ifrevision
\revisionfalse     %
\usepackage{xcolor}
\usepackage{multirow}

\definecolor{RED}{rgb}{1.0, 0, 0}

\usepackage{xeCJK}
\setCJKsansfont[Path=fonts/]{BabelCJKSans-Regular.otf}
\setCJKmonofont[Path=fonts/]{BabelCJKMono-Regular.otf}

\setcopyright{none}
\acmDOI{}
\acmISBN{}

\author{Zixiaofan Yang}
\affiliation{%
  \institution{Boston University}
  \city{Boston}
  \country{USA}
}

\author{Chang Xiao}
\affiliation{%
  \institution{Boston University}
  \city{Boston}
  \country{USA}
}

\hypersetup{pdfauthor={Zixiaofan Yang and Chang Xiao}}
\begin{document}
\renewcommand{\shortauthors}{Yang and Xiao}

\title[Deciphering the Babel of Play]{Deciphering the Babel of Play: A Human-AI Collaborative Approach for Large-Scale Cross-Language Analysis of Game Reviews}

\begin{abstract}
\input{sections/0_Abstract}

\end{abstract}



\maketitle

\input{sections/1_Introduction}

\input{sections/2_Related-Works}
\input{sections/3_Dataset}
\input{sections/5_LLM_Scoring}

\input{sections/6_Review_Analysis}

\input{sections/7_Limitation}

\input{sections/8_Conclusion}

\bibliographystyle{ACM-Reference-Format}
\bibliography{bib/steam}
\input{sections/appendix}

\end{document}

%% file: sections/0_Abstract.tex
We present a large-scale cross-language analysis of game reviews using a human-AI collaborative framework that combines quantitative screening with multilingual large language models (LLMs). Starting from 17 million Steam reviews across 30 languages and 2,000 top-selling titles, we select 28 games with notable cross-language rating patterns. We then apply LLM-assisted content analysis to 442,162 reviews spanning 17 languages, with human researchers guiding codebook development and interpreting the results.
Our findings reveal differences in both the aspects language communities prioritize and how they evaluate them, highlighting the roles of narrative expectations, game mechanics and stability, localization quality, cultural proximity, and perceptions of developers and publishers. We also identify rare cases of cross-language consensus. This work offers empirical insights into cross-cultural game evaluation and a scalable methodological approach to multilingual content analysis that preserves human interpretation.

%% file: sections/1_Introduction.tex
\section{Introduction}

When you play a game, what do you care about most? Is it the level of challenge, the art style, or the story? Do players around the world care about the same things you do?
In many digital media, culture shapes the ``lens'' through which quality is perceived and judged. Research in communication, cultural studies, and media has repeatedly shown that audiences from different countries evaluate the same artifact in systematically different ways~\cite{hall1976beyond, lee2008hollywood, taneja2016global, liebes1994export, hofstede2011dimensionalizing}. In Human-Computer Interaction (HCI), it is also documented that users from different language and regional groups apply different criteria when assessing interactive digital systems in terms of usability, aesthetics and satisfaction~\cite{reinecke2011improving, tractinsky1997aesthetics, marcus2000cultural, ge2024culture, ito1996impact}.

Video games are no exception. As a global medium and highly interactive digital artifact, games are discussed across dozens of languages and continually reinterpreted through local expectations and experiences. Because games are distributed and consumed primarily through digital platforms~\cite{app2top_gaming_industry_2024}, players routinely leave written assessments in their own languages on services such as Steam~\cite{steam}, generating large volumes of publicly accessible reviews. Although language is a rough proxy for cultural groups, reviews grouped by language offer a natural lens for comparing how communities value and critique the same game.

Prior research has identified cross-cultural patterns in both overall evaluations and specific priorities. East Asian players tend to give lower ratings than Western players~\cite{ma2025cultural}, while Japanese players place greater emphasis on bugs and localization and American players on replay value~\cite{zagal2013cultural}. Media coverage and online discussions also highlight cross-language differences arising for varied reasons. Warframe (2013) enjoyed a strong global reputation but saw its Steam score fall after a political controversy prompted negative reviews from Chinese players~\cite{talbot2019warframe_review_bomb}. Torchlight: Infinite (2022) received mixed English reviews criticizing pay-to-win elements, while Korean players, more accustomed to free-to-play monetization, responded more favorably ~\cite{turbostrider27_steam_language_review_updates_reddit}.
More recently, Hollow Knight: Silksong (2025) received widespread acclaim but much lower ratings in Simplified Chinese, where players criticized an overly flowery and imprecise localization for distorting the original meaning~\cite{gach2025reviewbomb_silksong}.

Understanding how different communities assess the same game benefits both researchers and developers. For researchers, game reviews offer a rich lens on how culture shapes play: what counts as fair or unfair, which designs are valued, and how players discuss privacy, monetization, and trust in developers and publishers. Comparing these assessments across languages provides an opportunity to revisit classic questions in cross-cultural HCI, including how design values travel, transform, or clash when interactive systems move across borders.
For developers, these insights can reveal where and why a game fails to resonate, which issues players consider severe or forgivable, and how localization interacts with cultural expectations. This can inform design and patching decisions, as well as localization strategies that go beyond literal translation to reflect players' values.

Despite this promise, cross-language analysis of game reviews remains technically and logistically challenging. Traditional qualitative methods require human coders with sufficient fluency to interpret nuance. Assembling teams capable of coding thousands of reviews across ten or more languages with native-level proficiency is difficult and costly. Moreover, the sheer volume of reviews for popular titles makes purely manual coding at scale infeasible. Classical natural language processing (NLP) and machine translation can help, but translation errors and inconsistent handling of gaming jargon, memes, and culture-specific idioms can obscure meaning. These challenges make it difficult to combine broad coverage across games and languages with detailed, interpretable analysis of players' assessments.
As a result, few large-scale studies systematically compare how different language communities assess the same game.

Recent advances in multilingual large language models (LLMs), including GPT~\cite{openai_gpt6_astra_system_card_2026} and Gemini~\cite{googledeepmind2026gemini31pro}, create a new opportunity to address this gap. Empirical evaluations demonstrate strong performance in multilingual reasoning~\cite{alshammari2026mathnet} and sentiment and emotion annotation across languages~\cite{rathje2024gpt}.
Furthermore, in qualitative research, LLMs can apply human-developed codebooks with agreement comparable to human coders~\cite{dutta2026mining} and produce coherent themes grounded in the underlying data~\cite{qiao2025thematic,wang2026centaurta}. Human--LLM collaboration also supports codebook refinement and the identification of additional concepts~\cite{meng2026exploring}, enabling LLMs to serve as scalable ``second coders'' under researcher guidance. 
Together, these developments suggest that LLMs can support multilingual coding at a scale that is difficult to achieve manually.

In this paper, we ask both a ``how'' and a ``what'' question. First, how can multilingual LLMs support large-scale cross-language analysis of game reviews while preserving human interpretation? We develop a human-AI collaborative framework that allocates work according to complementary strengths: the LLM performs labor-intensive coding and theme consolidation, while researchers guide codebook development, refine outputs, and retain responsibility for final interpretation. 
Second, what does this analysis reveal about similarities and differences in how language communities evaluate the same games? We examine which aspects players prioritize, what they praise or criticize, and where their evaluations converge or diverge.

We conduct our study on Steam, starting from 17 million reviews across 30 languages and 2,000 top-selling games. Through quantitative screening and manual selection, we identify 28 titles with notable cross-language rating patterns. We then apply our framework to 442,162 sampled reviews spanning 17 languages, comparing which aspects different language communities emphasize and how they evaluate them, with human researchers interpreting the findings in context.

Our contributions are threefold.
First, we develop a human-AI collaborative framework that enables large-scale cross-language review analysis by overcoming barriers posed by data volume and language differences. We provide full prompt templates to support replication and further research.
Second, we present an in-depth case study of the widely discussed \textit{Clair Obscur: Expedition 33}, revealing contrasting interpretations of narrative choices and differences in which aspects each language community prioritizes.
Third, we identify recurring cross-language patterns across the 28 selected titles involving localization quality, cultural proximity, game mechanics and stability, and perceptions of developers and publishers, alongside rare cases of cross-language consensus.
These contributions offer both a reusable methodological approach and empirical insights for cross-cultural game research.

%% file: sections/2_Related-Works.tex
\section{Background}

\subsection{Understanding Player Experience via Game Reviews}

Game reviews provide a valuable window into how players experience, interpret, and evaluate games, making them an important resource for understanding player behavior and informing game user research. Players have shared assessments on game-specific websites, from regional platforms such as Famitsu~\cite{famitsu} and A9VG~\cite{a9vg} to Metacritic~\cite{metacritic} and GameSpot~\cite{gamespot}, as well as general marketplaces like Amazon. With the growth of Steam~\cite{steam}, these reviews have become globally accessible, allowing researchers to observe player experiences at unprecedented scale.

The research community has studied game reviews from multiple perspectives to understand player experience. Early studies such as Sirbu et al.~\cite{sirbu2016extracting} used sentiment analysis to capture affective signals, illustrating how players communicate emotional reactions. Content-focused analyses further highlight the experiences players choose to express in their own words. Ahn et al.~\cite{ahn2017makes} identified factors of game experience that make a game successful among players. Lin et al.~\cite{lin2019empirical} manually coded reviews into categories including pro, con, suggestion, and bug, demonstrating the range of experiential cues reviews provide; hybrid qualitative and topic-modeling approaches further examine reviews' contextual meanings~\cite{tong2025game}. Computational topic modeling extracts thematic components of player experience from large review corpora, revealing which dimensions players collectively emphasize, such as narrative and achievement~\cite{wang2020components,pielka2019community,busurkina2020game}, while related content analyses identify environmental themes for ecogame classification~\cite{edlinger2025analyzing}.

Researchers have also used reviews to investigate broader contexts shaping player experience. Prior work examines how the COVID-19 pandemic affects emotional expression~\cite{petrosino2022panorama}, how online achievements shape engagement~\cite{rizani2023application}, how review length and emotion differ across genres~\cite{guzsvinecz2023length}, how parents express concerns about mobile games~\cite{winter2025leave}, and how experience and enjoyment vary over time and across platforms~\cite{dutta2026mining,abdelqader2026using}. Other studies connect reviews to commercial impact by comparing how user and critic scores influence sales~\cite{kraemer2025all}. A complementary line of research asks what makes reviews helpful for other players, identifying meta-information and text features associated with perceived helpfulness and funniness~\cite{kang2017study,wang2021explaining}.

Beyond user-generated review corpora, research has also investigated player preferences using surveys and theoretical frameworks. Tondello et al.~\cite{tondello2017framework} proposed a preference framework linking demographic factors to favored game elements and play styles, while Vahlo et al.~\cite{vahlo2017digital} combined review articles and population-level surveys to identify game dynamics and player types. Subsequent work introduced trait-based models that characterize player preferences and complement review-based understandings of player experience~\cite{fortes2018towards}.

Despite this rich body of work, research on game reviews and player preferences remains largely centered on English-speaking populations. Studies that include multilingual reviews often compare games or platforms without examining differences between language communities. This limited attention to cross-cultural variation constrains our understanding of global player experiences.

\subsection{Cross-Cultural Game Studies and Localization}

Prior cross-cultural work on video games typically trades breadth in games and cultures for depth, focusing on small samples or single titles and offering higher analytical granularity at the cost of representativeness.

Within this space, studies of review content provide clear evidence of cultural variation in how players evaluate games. \citet{zagal2013cultural} compared U.S. and Japanese reviews, finding greater sensitivity to bugs and localization among Japanese players and greater emphasis on replay value among American players. \citet{bruckner2019exploring} examined three JRPGs in Germany and Japan through qualitative analysis of media articles and user reviews, using bottom-up open coding to identify nine categories such as Audio-Visual or Rules/Mechanics. Their subsequent think-aloud study linked cultural background to audio-visual and narrative experiences, with little effect on mechanics~\cite{bruckner2020play}. Recent comparisons extend to 17 Souls-like games~\cite{pan2024cultural} and four culturally representative games~\cite{koyama2025role}, but each covers only three languages; detailed thematic analysis of localization reception similarly covers three languages for one title~\cite{hsu2025international}. At a broader platform level, \citet{ma2025cultural} found lower ratings in East Asian language communities across 510 games, while their text analysis covered 1,950 Chinese and English reviews from 39 games. 

Beyond reviews, other cross-cultural studies broaden the lens to play behavior, identity, and player motivation. Prior work compares pathways to gaming disorder and play motivations across limited national pairs, examines gamer identity and cultural orientations in specific populations, and meta-analyzes Eastern vs.\ Western responses to violent games~\cite{chew2024cross,kahn2015trojan,cwil2020cross,anderson2010violent,lee2012there}. Developmental and behavioral studies similarly contrast play patterns and telemetry across a small set of regions or age groups~\cite{colwell2005video,bialas2014cultural,koban2020further}. Other studies examine localization preferences, representation, culturally adapted designs, and L2 learning benefits in specific titles or subgroups~\cite{yu2025cultural,novak2024cross,shaw2010identity,pyae2018understanding,soyoof2018video}.

Finally, much scholarship on game localization takes a conceptual or theoretical orientation, examining the balance between cultural fidelity and adaptation~\cite{pelletier2011video}, culturalization beyond translation that accounts for geopolitical norms and sensitivities~\cite{edwards2011culturalization}, and localization's role in shaping cultural imaginaries~\cite{carlson2011imagined}. Related work addresses how values and audience norms guide market-specific adaptation~\cite{i2016games} and dimensions of localization quality~\cite{bernal2020key}, while empirical trade analysis examines localization's economic value~\cite{yang2025tower}. Broader surveys and handbooks likewise call for deeper cultural treatment beyond linguistics~\cite{chandler2011game,o2013game,o2015game,mangiron2017research,bernal2018creativity}.

\subsection{LLM-Assisted Content Analysis}

Recent advances in LLMs open new paths for large-scale online content analysis. Their multilingual understanding and natural-language outputs support coding, thematic analysis, and summarization of qualitative data~\cite{zhu2024multilingual,ziems2024can}. In game user research, Dutta et al.~\cite{dutta2026mining} combine human inductive codebook development with LLM deductive coding of game reviews, reporting human--LLM agreement comparable to human--human agreement. Earlier evaluations demonstrated fair-to-substantial agreement in codebook-based deductive coding~\cite{xiao2023supporting}, with subsequent work improving agreement through codebook and prompt refinement~\cite{dunivin2025scaling}. For inductive coding, Qiao et al.~\cite{qiao2025generative} reported that human reviewers accepted approximately 85\% of LLM-generated codes. More recently, Hill et al.~\cite{hill2026large} found that GPT-5 met non-inferiority criteria for deductive agreement and expert-rated inductive analysis of a healthcare focus-group transcript. Beyond coding, studies demonstrate nuanced theme identification~\cite{byun2023dispensing}, thematic summaries aligned with human interpretations~\cite{castellanos2025large}, and more interpretable, human-aligned topic discovery than traditional models~\cite{pham2024topicgpt}.

Given these advances, researchers have developed automatic analysis pipelines~\cite{parfenova2024automating,than2025updating,qiao2025thematic} and human--AI collaborative tools such as CollabCoder, SenseMate, PaTAT, and DeTAILS~\cite{gao2024collabcoder,overney2024sensemate,gebreegziabher2023patat,sharma2025details}. Applications span inductive coding and concept development~\cite{gao2024collabcoder,qiao2025thematic,wang2025lata,de2024performing,lam2024concept,hamalainen2023evaluating,zhong2025hicode}, deductive coding and code suggestion~\cite{hila2025assessing,spinoso2023qualitative,xiao2023supporting,chew2023llm}, validation against human annotations~\cite{pangakis2025keeping}, and iterative code consolidation and theme construction with human input~\cite{meng2026exploring,wang2026centaurta,bai2026multimodule,bang2026developing}. Together, they provide precedents for integrating LLM assistance across analytical stages while retaining researcher involvement.

Evidence also supports cross-language analysis: psychological text annotation across 12 languages~\cite{rathje2024gpt}, Spanish news annotation outperforming outsourced coders against expert labels~\cite{bermejo2025llms}, accurate annotation of predefined linguistic features in Serbian headlines~\cite{bago2026corpus}, and accurate sentiment classification with human-aligned explanations for code-mixed Swahili, Sheng, and English~\cite{ochieng2026reasoning}. These findings strengthen the case for LLM-assisted multilingual review analysis.

At the same time, LLMs remain debated instruments in the qualitative research community because of concerns about bias and validity~\cite{jiang2021supporting,schroeder2025large}. To make use of the complementary strengths of humans and LLMs, we design our human--LLM system to draw on capabilities with accumulating evidence of reliability, such as text summarization (coding individual quotes), multilingual understanding, and few-shot learning from human examples, while leaving the higher-level tasks of guiding the analysis and interpreting its results to human researchers.

%% file: sections/3_Dataset.tex
\section{Selecting Games with Cultural Differences}\label{sec:select}

\subsection{Initial Dataset}

To study player opinions across countries, we use Steam, the largest PC game distribution platform~\cite{rokky_pc_distribution_2025}, available in more than 200 countries and regions. 
Steam allows users to select a language when posting a review and supports retrieving reviews by language. We treat this user-selected language as a proxy for cultural community: for example, users who select Simplified Chinese are likely to speak Chinese, follow Chinese-language media, and share Chinese cultural references. We therefore compare language-specific review sets to examine how these communities evaluate the same game.

We begin with the 2,000 top-selling games on Steam, obtained through the Steam Store API using its top-sellers ranking. Beyond this cutoff, review counts drop sharply, often falling below 10 in non-English languages and providing limited data for cross-language comparison. We therefore expect these 2,000 games to include essentially all titles with sufficient review volume for meaningful analysis of cultural differences.
We collected all reviews of these games using the Steam Reviews API~\cite{woctezuma_steamreviews}, including review text, binary ratings, update timestamps, and language.

Because patches and content updates can change player perceptions over time, we restrict our initial dataset to reviews posted between August 1, 2024 and August 1, 2025 to reduce temporal confounds. This one-year window retains sufficient review volume for cross-language comparison. Applying this filter yields 17 million reviews in 30 languages. The most common review language is English (7.63M), followed by Simplified Chinese (2.68M), Russian (2.24M), Brazilian Portuguese (816K), European Spanish (738K), German (510K), Turkish (403K), French (397K), Polish (350K), and Korean (288K).

\subsection{Finalizing the Set of Games for Study}

Although we could apply the full pipeline to all games, our initial observations showed that the majority exhibit little or no systematic cross-language differences in player evaluations. We therefore use quantitative screening to select a focused set of games with notable cross-language rating patterns for in-depth analysis.

We screen the initial dataset using three selection criteria. First, we examine variation in each game's ratings across languages to identify unusually high disagreement (Cross-Language Disagreement). Second, we use the same measures to identify games with low variation and consistent evaluations across languages (Cross-Language Agreement). Third, we compare each game's rating-based rank within a language with its mean rank across languages, identifying titles unusually favored or disfavored in a particular language (Ranking Outlier). The formal definitions of these metrics are provided in Appendix~\ref{app:selection}.
We rank games under each selection criterion and select the top-ranked cases for qualitative analysis. While screening determines the games examined in this study, our human-AI collaborative content analysis framework can also be applied directly to games of interest without this step.

In addition to the quantitatively selected titles, we manually include three highly popular games highlighted in media coverage and online discussions for their cross-language rating differences (Manual Selection). One of these, \textit{Hollow Knight: Silksong}, was added after the collection cutoff. Table~\ref{tab:games_languages} summarizes the 28 selected games and languages analyzed.

\begin{table*}[t]
\centering
\begin{tabular}{l l c}
\hline
\textbf{Game Name} & \textbf{Selected Languages} & \textbf{Selection Criterion} \\
\hline

TEKKEN 7 & Brazilian, Chinese (S) &
    \multirow{12}{*}{\parbox{3cm}{\centering Cross-Language Disagreement}} \\
Borderlands 2 & Italian, Spanish, Chinese (S) & \\
EA SPORTS FC 25 & Indonesian, Japanese & \\
SCUM & Brazilian, Korean & \\
Monster Hunter Wilds & Vietnamese, German, Chinese (S) & \\
Sun Haven & Brazilian, Chinese (T) & \\
S.T.A.L.K.E.R.: Shadow of Chornobyl & Ukrainian, Russian & \\
Dead by Daylight & Romanian, Brazilian, Chinese (S) & \\
GUILTY GEAR -STRIVE- & Spanish, Chinese (S) & \\
Call of Duty: Modern Warfare & Russian, French & \\
Starfield & Brazilian, Korean & \\
HELLDIVERS 2 & Brazilian, Chinese (S) & \\
\hline\hline

RimWorld & Top 10 languages &
    \multirow{4}{*}{\parbox{3cm}{\centering Cross-Language Agreement}} \\
Balatro & Top 10 languages & \\
Undertale & Top 10 languages & \\
DELTARUNE & Top 10 languages & \\
\hline\hline

Black Myth: Wukong & Chinese (S), Other 9 languages &
    \multirow{9}{*}{\parbox{3cm}{\centering Ranking Outlier}} \\
Fast Food Simulator & Korean, Other 9 languages  & \\
Metro Exodus & Russian, Other 9 languages  & \\
Deep Rock Galactic: Survivor & Turkish, Other 9 languages  & \\
Blasphemous & Spanish, Other 9 languages  & \\
Age of Empires II & Korean, Other 9 languages  & \\
God of War & Chinese (S), Other 9 languages  & \\
Satisfactory & Turkish, Other 9 languages  & \\
Frostpunk & German, Other 9 languages & \\
\hline\hline

Clair Obscur: Expedition 33 & Chinese (S), English &
    \multirow{3}{*}{\parbox{3cm}{\centering Manual Selection}} \\
The Last of Us Part II Remastered & Chinese (S), English & \\
Hollow Knight: Silksong & Chinese (S), Other 9 languages & \\
\hline

\end{tabular}
\caption{Selected games, languages, and selection criteria. ``Brazilian'' indicates Brazilian Portuguese; ``Chinese (S)'' indicates Simplified Chinese; ``Chinese (T)'' indicates Traditional Chinese; and ``Spanish'' indicates European Spanish. ``Top 10'' denotes the ten most common review languages in our dataset; ``Other 9'' excludes the focal language.}
\label{tab:games_languages}
\end{table*}

%% file: sections/5_LLM_Scoring.tex
\section{Human-AI Collaborative Content Analysis}\label{sec:llm}

Content analysis of user reviews is both context dependent and methodologically demanding: game reviews are highly nuanced and still require human researchers as the final judges of meaning. At the same time, the size and language diversity of our corpus make purely manual coding infeasible. We therefore design a human-AI collaborative workflow that allocates work according to complementary strengths: the LLM performs fine-grained, labor-intensive operations such as multilingual coding and sentiment annotation at the level of individual reviews, while human researchers lead codebook design, supervise code merging into themes, and interpret the resulting patterns.

\subsection{Method}

Since we aim to discover themes from the data rather than impose an \textit{a priori} schema, we employ an \textit{Exploratory Sequential Design}~\cite{ford2004content, drisko2016content}. This two-phase mixed-methods design first establishes a qualitative understanding of the data to build a valid instrument (the codebook), which is then applied quantitatively.

\begin{figure*}[t]
  \centering
  \includegraphics[width=0.8\linewidth]{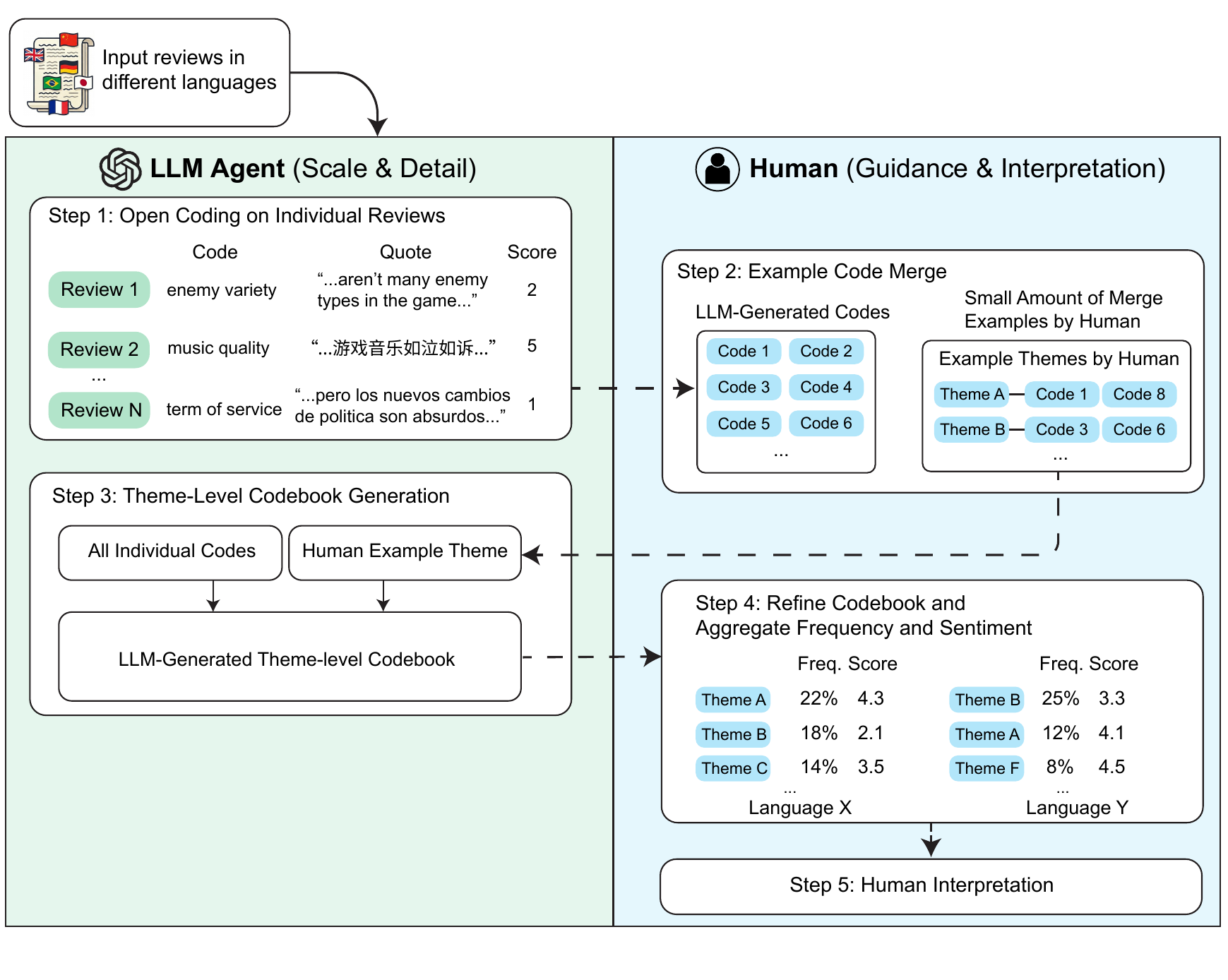}
  \caption{Overview of our human-AI collaborative content analysis workflow.}
  \label{fig:llm_workflow}
\end{figure*}

The LLM first conducts inductive open coding on individual reviews, producing \textbf{initial codes} in English with supporting quotes in the original languages. Human researchers then review the initial codes and provide examples of the desired level of abstraction for consolidating these codes into \textbf{themes}. Next, the LLM iteratively assigns codes and proposes new themes to produce a \textbf{theme-level codebook}. Finally, human researchers inspect and refine the codebook, adjust theme assignments, and interpret the aggregated theme frequencies and sentiment scores.
Figure~\ref{fig:llm_workflow} shows an overview of the workflow. For each game, it produces a unified theme-level codebook across languages, with each theme's frequency and average sentiment calculated separately for each language.

\subsubsection{LLM Configuration}
Unless otherwise specified, we use Gemini 3.1 Pro model~\cite{googledeepmind2026gemini31pro}, a strong reasoning model that demonstrates leading performance on multilingual benchmarks and achieves accuracy comparable to human experts~\cite{alshammari2026mathnet,rein2024gpqa}. To support reproducibility and control the prompts, we access the model through the Gemini API rather than the chatbot interface, whose internal system prompts are less transparent and not user configurable. All parameters are set to their default values.

\subsubsection{Individual Review Analysis}

For each call, the LLM receives a single review and produces initial codes with neutral English names, each capturing a distinct aspect of game assessment.
To ground each code in the review text, we ask the LLM to (i) extract verbatim evidence quotes in the original language, (ii) provide English translations of non-English quotes, and (iii) explain how the quotes support the code.
The LLM also assigns a sentiment score indicating the attitude expressed toward each coded aspect, from 1 (very negative) to 5 (very positive). A score of 3 indicates neutral or mixed sentiment and is also used when clear sentiment cues are absent. Appendix~\ref{app:prompt} provides the prompt for this coding process, and Table~\ref{tab:llm_codes_example} shows example outputs.

In our early exploration, randomly sampling 10,000 reviews per game per language produced approximately 1,000 \emph{lexically distinct codes}. Adding more reviews yielded diminishing returns in code diversity. We therefore adopted a sampling limit of up to 10,000 randomly selected reviews per game per language for computational and cost efficiency. This yields 442,162 reviews for LLM coding.

\begin{table*}[t]
\centering

\begin{tabular}{l  p{4.5cm} p{4.5cm} l}
\toprule
\textbf{Code}  & \textbf{Quote (Original Language)} & \textbf{Explanation} & \textbf{Sentiment} \\
\midrule
Level Variety & ``I love the fact that you get to experience different terrain and different stories in each of the locations.'' (English) & Environmental and narrative differences across locations are praised as a source of enjoyment. & 5 \\
\hline
VPN Policy  & ``Puoi essere bandito per l'utilizzo di una VPN durante il collegamento ai server online.'' (Italian, ``You can be banned for using a VPN while connecting to online servers.'') & Criticism centers on the risk of losing online access because of VPN use. & 1 \\
\hline
Character Variety  & ``...elegir entre varios personajes con habilidades únicas'' (Spanish, ``...choose between several characters with unique abilities'') & The range of characters and their distinct abilities provides appealing gameplay choices. & 4 \\
\hline
Language Support  & ``türkçe destegini kaldırmiş bende desteyını kaldırıyorum.'' (Turkish, ``They removed Turkish support so I am removing my support.'') & Withdrawing support expresses dissatisfaction with the removal of a language option. & 1 \\
\hline
Narrative Quality  &
    ``编剧应该给其他所有部门嗑一个''
    (Simplified Chinese, ``The screenwriter should kowtow to all other departments'') & The comparison with other departments presents the writing as a weakness in the game. & 1 \\
\hline
NPC Behavior  & 
``직원이 직접 춤까지 춰준다니까''
(Korean, ``The staff even dances for you personally'') & The staff NPCs' playful behavior is described with enthusiasm. & 5 \\
\hline
Stability  & 
``バグだらけなのはともかくとして''
(Japanese, ``Leaving aside the fact that it's full of bugs'') & The reference to widespread bugs identifies technical reliability as a problem. & 2 \\
\bottomrule
\end{tabular}
\caption{Examples of LLM-generated initial codes with supporting evidence, explanations, and sentiment scores (1 = very negative, 5 = very positive).}
\label{tab:llm_codes_example}
\end{table*}

\subsubsection{Theme-level Codebook Generation}

Although the initial codes are grounded in individual reviews, they are generated independently without global context. This results in lexical redundancy, where semantically identical concepts receive different code names (e.g., ``music quality'' vs. ``soundtrack''; ``replayability'' vs. ``replay value''). Analyzing these lexically redundant codes separately produces sparse signals that obscure meaningful patterns. To address this, we implement a \emph{Programming-by-Demonstration} approach to consolidate these codes into a concise theme-level codebook.

\paragraph{Exemplar Codebook Construction}

First, for each game, we aggregate all initial codes from all target languages into a single pool, grouping occurrences with identical code names. The authors then review the pooled codes alongside their quotes and explanations to merge them into broader themes.

Given prior research showing that LLMs are strong few-shot learners~\cite{brown2020language}, the human coders conduct an \emph{illustrative merge} to generate examples rather than manually merging all initial codes. This illustrative stage concludes when a coder identifies 10 distinct themes or has scanned all initial codes, whichever comes first. We empirically found that 10 theme examples are sufficient for the LLM to learn the appropriate level of abstraction.

To ensure consistency during codebook construction, the coders follow three guiding principles:
\begin{enumerate}
\item \textbf{Semantic Equivalence:} Merge codes only when they have similar meanings within the gaming context (e.g., merging ``soundtrack'' and ``background music'' into ``Music''). Related but distinct concepts, such as ``Voice Acting'', must remain separate to preserve specific feedback nuances.

\item \textbf{Mutual Exclusivity:} Themes must be distinct and non-overlapping to prevent ambiguity.

\item \textbf{Contextual Fidelity:} The merged theme must accurately reflect the specific aspect described in the original evidence quotes, rather than relying on the code name alone.
\end{enumerate}

\paragraph{LLM-Based Codebook Expansion}

Starting from this small-scale, human-created exemplar codebook, we use the LLM to expand it through alternating steps of code assignment and theme proposal.
In each assignment call, the LLM receives the current codebook and a list of unassigned initial codes, accompanied by representative evidence quotes and their corresponding explanations. It assigns each code to exactly one existing theme when there is a clear match; otherwise, the code is placed in an ``uncategorized'' list.
After a full assignment pass, the LLM receives the remaining uncategorized codes and the current codebook. Following the same three merging principles and using the established themes to guide abstraction and naming, it proposes new, non-overlapping themes. 
We add the new themes and their grouped initial codes to the codebook, then repeat assignment and theme proposal for the remaining codes until all initial codes have been assigned to a theme. The detailed prompts for both steps are provided in Appendix~\ref{app:prompt}.

\paragraph{Human Refinement and Aggregation of Theme Statistics}

Finally, the two authors jointly inspect the codebook and discuss whether to reassign codes or split themes. In most cases, the codebook is already coherent and consistent with the exemplar codebook's structure and granularity, requiring only occasional small adjustments.

After refinement, each unique code name is linked to exactly one theme. We retain all code occurrences with their language, quotes, explanations, and sentiment scores, allowing us to trace themes back to individual reviews.

For each game and language, we calculate theme frequencies and average sentiment. Theme frequency is the percentage of analyzed reviews containing the theme, counting each review once and including reviews yielding no codes in the denominator.
We interpret theme frequency as a proxy for the relative importance of aspects in players' expressed evaluations~\cite{mirzababaei2020postmortem,lin2019empirical}, assuming that reviewers tend to foreground important or strongly perceived aspects~\cite{dutta2026mining}.
We calculate average sentiment by first averaging the theme's code scores within each review, then averaging across reviews containing it. When grouping languages together, we report macro-averages of language-level theme frequencies and sentiment scores.
These summaries support final human interpretation by showing which aspects each language community emphasizes and how positively or negatively it evaluates them.

\subsection{Validation of LLM-Assisted Coding}

Before applying the workflow to all games and languages, we conducted a small-scale human validation to assess its suitability for our task.

To validate the LLM-generated initial codes, we randomly sampled 200 English and 200 Chinese reviews of \textit{Clair Obscur: Expedition 33}. Two authors proficient in both languages and familiar with the game evaluated the outputs, which included 558 supporting quotes and 167 lexically distinct initial codes.
The authors checked each quote against its source review and judged whether its associated code accurately reflected the quoted content in the game's context. All 558 quotes appeared in their source reviews, with no fabricated quotes observed. Both authors judged 531 code--quote pairs (95.2\%) accurate, disagreed on 18 (3.2\%), and judged 9 (1.6\%) inaccurate. Errors mainly involved vague comments whose assessed aspect was unclear.
The authors also assigned sentiment scores using the same 5-point scale. The LLM and both human coders agreed on sentiment polarity (positive, negative, or neutral/mixed) in 96.6\% of cases. Agreement on the 5-point scale was also good (ICC$(2,1) = 0.77$), comparable to agreement between the human coders (ICC$(2,1) = 0.79$). Given the subjectivity of fine-grained sentiment judgments, we consider this agreement sufficient for our analyses.

For code-to-theme assignment, we randomly sampled 100 assignment decisions across iterations on \textit{Clair Obscur: Expedition 33}, including ``uncategorized'' decisions. Using the same codebook and evidence available to the LLM at each iteration, the authors assigned each code without seeing the model's decision. The LLM and human assignments matched in 95\% of cases.
For new theme proposals, the authors jointly evaluated 50 proposed themes by examining each theme’s name, grouped codes, and supporting evidence alongside the existing codebook. Each theme was marked as acceptable or requiring changes according to the three merging principles: semantic equivalence, mutual exclusivity, and contextual fidelity. 92\% of themes were accepted without revision.

This small-scale validation provides an initial estimate of agreement between the LLM and human judgments; a comprehensive evaluation across all games and languages is beyond the scope of this paper. Human inspection and refinement of the codebook provide an additional check on coding errors before interpreting aggregate theme frequencies and sentiment patterns~\cite{ziems2024can,pangakis2025keeping}. Together with existing benchmarks of LLMs' multilingual language understanding~\cite{singh2025global,chang2025global,googledeepmind2026gemini31pro}, these results provide further evidence that our prompting strategy and human-AI workflow can adequately support the multilingual coding and codebook expansion required in this task.

%% file: sections/6_Review_Analysis.tex
\section{Results and Discussion}

In this section, we first present a case study with detailed coding results to showcase the potential of our framework for in-depth qualitative analysis. We then summarize broader findings that recur across games and language communities and can inform future studies of cross-cultural game appreciation.

\begin{figure}[t]
  \centering
  \includegraphics[width=1.0\linewidth]{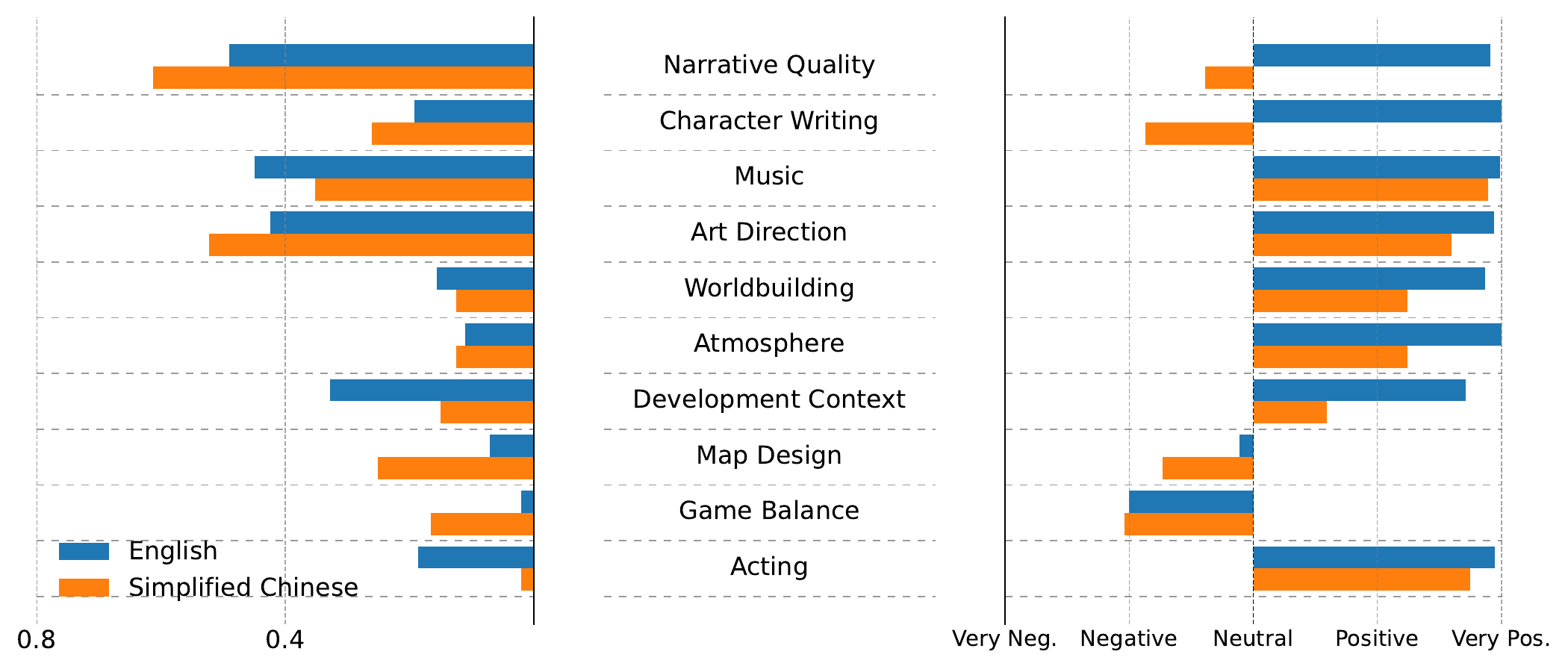}
  \caption{Theme frequencies (left) and average sentiment scores (right) for English and Simplified Chinese reviews of \textit{Clair Obscur: Expedition 33}.}
  \label{fig:33}
\end{figure}

\subsection{Case Study on \textit{Clair Obscur: Expedition 33}}

The widely acclaimed game \textit{Clair Obscur: Expedition 33} has been extensively discussed in English- and Chinese-speaking communities, with media reports and online forums frequently highlighting differences in opinion. To understand this divergence, we apply our pipeline to the game's English and Simplified Chinese reviews. The resulting theme frequencies and sentiment scores are shown in Figure~\ref{fig:33}.

First of all, the most frequently mentioned theme, \textbf{Narrative Quality}, shows a substantial sentiment gap: English reviews average an almost perfect 4.91, compared to 2.61 in Simplified Chinese, leaning negative. We examined this theme’s initial codes, explanations, and quotes to understand this gap.
Simplified Chinese reviews emphasize plot structure and logic, with recurring codes such as narrative clarity, coherence, and consistency (example excerpt: ``格局从一开始宏大的的远征救世逐渐变小到家庭伦理剧，前后割裂感严重'', ``The story's scope gradually shrinks from a grand expedition to save the world to a family drama, with a severe disconnect between the earlier and later parts''). English reviews instead emphasize emotional investment and an absorbing narrative arc, reflected in codes such as narrative impact and immersion (example excerpt: ``Ending was a rough pull on the heartstrings'').
These differing priorities help explain the contrast between structural criticism in Simplified Chinese reviews and appreciation of the story's emotional impact in English reviews.
Much of this divergence centers on the story's pivot from an apparent world-saving quest to one family's grief and unresolved emotions. Many English-speaking players find this focus on personal loss and psychological healing thematically rich and artistically meaningful. By contrast, Chinese-speaking players often expect clear collective stakes and a shared world worth saving, making the narrower family focus feel less consequential. This contrast may reflect an individualist emphasis on inner experience in English-language reviews and collectivist expectations of shared stakes in Simplified Chinese reviews.

\textbf{Character Writing} shows an even larger sentiment gap, scoring a perfect 5.00 in English reviews and 2.13 in Simplified Chinese. Within this theme, Simplified Chinese reviews criticize the abrupt protagonist switch for disrupting identification with the playable character (example excerpt: ``你让我如何代入'', ``how am I supposed to identify with him?''). English reviews instead praise the depth of characterization and the emotional connections developed through character interactions (example excerpt: ``memorable characters with emotional complexity'').

Comparing the frequencies, we also see an interesting pattern. Despite lower sentiment, Simplified Chinese reviews discuss \textbf{Narrative Quality} and \textbf{Character Writing} more frequently, at 61.31\% and 26.13\%, compared to 48.99\% and 19.19\% in English. This suggests their criticism reflects active engagement and a desire to share concerns and interpretations, rather than indifference.
This interpretation draws on \citet{cui2018social}'s broader discussion of how unmet expectations for participation can contribute to negative social evaluations. In our case, reviews may provide a visible channel for players to express their concerns to other players and developers.

On the other hand, both language communities praise the game's audiovisual elements. \textbf{Music}, \textbf{Art Direction}, \textbf{Worldbuilding}, and \textbf{Atmosphere} each appear in over 10\% of reviews in both languages, with average sentiment scores above 4.00. Players across languages highlight the painterly art style, richly textured environments, layered soundtrack, and cohesive worldbuilding that ties these sensory elements together. Although these aspects are generally praised, Chinese reviews have slightly lower average sentiment scores, consistent with previous research on cultural differences in rating styles and average evaluations~\citep{chen1995response,hong2016culture,ma2024cultural}.

Turning to theme frequencies, we also find differences in what players prioritize despite shared sentiment polarity. 

One notable example is \textbf{Development Context}, which captures perceptions of the studio and publisher, including their skill and resources, as well as reflections on passion, ethics, and working conditions. Both groups express positive sentiment, scoring 4.71 in English and 3.59 in Simplified Chinese. Yet the theme appears in 32.83\% of English reviews compared to 15.08\% in Simplified Chinese. This suggests that English players place more weight on industry and studio narratives, while Chinese players focus more on the playable experience.
\textbf{Acting}, covering character acting and voice performance, also appears much more frequently in English reviews (18.69\%) than in Simplified Chinese reviews (2.01\%), suggesting that performance receives greater attention in English-language discussions.

The frequency pattern reverses for \textbf{Map Design}, covering map guidance and navigation, and \textbf{Game Balance}, covering numerical balance in combat and character builds. Both groups express negative sentiment, but Chinese reviews mention these themes far more frequently, at 25.13\% and 16.58\%, compared to 7.07\% and 2.02\% in English. This indicates that Chinese players place greater emphasis on moment-to-moment gameplay, including navigation and balance tuning.

Overall, this case suggests that widely acclaimed storytelling can resonate differently across language communities as players bring different expectations to the game. Our method highlights where Chinese and English players differ, where they agree, and what each group prioritizes, enabling targeted, interpretable analysis of cultural differences in game experience.

\subsection{Findings across All Selected Games}

We further analyze all games in Table~\ref{tab:games_languages} and summarize the major findings below.
The most frequently discussed themes, their frequencies, and sentiment scores are available in Appendix~\ref{app:results}.

\subsubsection{Localization Matters}

As media coverage has reported~\cite{gach2025reviewbomb_silksong}, \textit{Hollow Knight: Silksong} was heavily review-bombed in Simplified Chinese over poor localization, with a Steam rating of 41.14 compared to an average of 92.68 across the other nine languages. Our human-AI collaborative content analysis identifies localization as the central concern. In the pooled baseline of reviews from the other nine languages, sentiment is generally positive toward character presentation, art direction, combat mechanics, and audio, and only slightly negative toward difficulty. Simplified Chinese reviews, however, tell a sharply different story: localization is the most frequently discussed theme, alongside criticism of difficulty, level design, and other gameplay systems. This divergence illustrates how translation quality can reshape the broader evaluation of a game.

\textit{Sun Haven} provides a complementary example, with multiple languages affected by incomplete, low-quality community translations. Traditional Chinese reviews give the game a low Steam rating of 28.44, with localization quality and developer priorities emerging as top themes, both receiving very negative sentiment. By contrast, Brazilian Portuguese reviews give the highest rating of 86.67, yet still frequently discuss localization quality. This suggests that localization problems carry different weights in players' overall recommendations across cultures.

Players also use reviews to directly criticize missing localization support. In \textit{SCUM}, Korean players primarily complain about the removal of promised Korean support, making localization and developer trust dominant themes. The Steam rating is only 28.81 in Korean, compared to 88.15 in Brazilian Portuguese. Korean reviews of \textit{Starfield} also frequently criticize missing localization. Similarly, \textit{Satisfactory}'s 1.0 launch lacked previously available Turkish localization, making localization and developer relations dominant themes in Turkish reviews.

\subsubsection{Cultural Proximity Enhances Game Reception}

Across several titles, players who share a game's cultural background respond especially positively.

\textit{Blasphemous} receives exceptionally high ratings among Spanish players. Drawing on southern Spain's religious art and traditions, especially Catholic Holy Week processions, it uses imagery these players readily recognize and value. Compared to the pooled baseline, Spanish reviews express stronger positive sentiment across themes, including visual art, narrative, atmosphere, and game mechanics.

A similar pattern appears in \textit{Metro Exodus}. Although highly rated globally, Russian reviews show a deeper emotional connection and more positive sentiment across most themes. Grounded in Russian literature and set in post-apocalyptic Russia, the game resonates strongly with audiences who share its cultural references.

The same dynamic appears in \textit{Black Myth: Wukong}, which receives an unusually high rating in Simplified Chinese. Many players celebrate it as a landmark Chinese-made blockbuster bringing their mythology to a global audience. Simplified Chinese reviews more frequently highlight culture, developer reception, and industry significance, with very positive sentiment.

These cases suggest that cultural familiarity amplifies a game's impact through meanings less accessible to international audiences, supporting previous work~\cite{pyae2018understanding} with large-scale quantitative evidence.

\subsubsection{Divergent Expectations of Narrative and Character Writing}

The divergence in narrative expectations identified in our \textit{Clair Obscur: Expedition 33} case study also appears in \textit{The Last of Us Part II Remastered}. Simplified Chinese reviews discuss narrative, themes, and characters more frequently and negatively than English reviews. Our coded reviews suggest that Chinese criticism centers on the shift to Abby's perspective and its disruption of attachment to the original protagonists, whereas English reviews more often praise the narrative direction as bold or subversive. Both communities nevertheless praise gameplay and graphics. This gap suggests that expectations about character attachment and empathy can shape how the same narrative choices are evaluated across language communities.

\subsubsection{Cross-Cultural Sentiment Gaps on Shared Game Aspects}

Reviews from different language communities may discuss similar aspects but express different sentiment, with East Asian communities generally more negative.

For \textit{EA Sports FC 25}, Indonesian and Japanese reviews identify similar problems with gameplay mechanics, controls, and AI behaviors, yet Japanese players express much stronger negativity. Similarly, Korean reviews of \textit{Starfield} are more negative than Brazilian Portuguese reviews on comparable topics. In \textit{Monster Hunter Wilds}, Simplified Chinese players discuss almost the same themes as German and Vietnamese players but express more negative sentiment toward performance, stability, content structure, and narrative.

These patterns are consistent with \citet{ma2024cultural}, who report systematically lower Steam ratings among East Asian players. Our analysis examines sentiment differences at the level of specific game aspects.

\subsubsection{Divergent Sensitivities to Stability and Mechanics}

Prior work identifies Japanese players as especially sensitive to bugs~\cite{zagal2013cultural}; our findings extend this observation to other East Asian player groups.
\textit{Age of Empires II: Definitive Edition} illustrates this difference. Reviews from the pooled baseline celebrate the game's longevity, visuals, and nostalgia, whereas Korean reviews are dominated by negative sentiment toward technical reliability, bug-fix speed, and developer relations.
A related pattern appears in \textit{GUILTY GEAR -STRIVE-}. Simplified Chinese players complain more frequently and intensely about game balance and combat mechanics, while Spanish players are broadly positive. Similarly, Simplified Chinese reviews of \textit{God of War} are more negative toward puzzle, enemy, and boss design, while other players focus more on narrative and visuals. This emphasis on mechanics echoes \citet{pan2024cultural}, who found greater attention to combat design in Chinese and Russian Souls-like reviews than in English reviews.
These comparisons complement prior findings on audiovisual and narrative experiences~\cite{bruckner2020play}, showing that cross-cultural differences also involve how players prioritize and evaluate game mechanics.

\subsubsection{Divergent Sensitivities to Privacy Concerns}

Language communities differ markedly in how privacy concerns influence their overall game assessments. \textit{Borderlands 2} is one of the most extreme cases of cross-language divergence in our dataset. Italian players focus mainly on traditional strengths, including characters, franchise perception, narrative, art direction, and co-op. By contrast, Spanish and Simplified Chinese reviews are dominated by very negative sentiment toward privacy and terms of service, followed by account enforcement, modding restrictions, and corporate ethics. This backlash followed Take-Two's February 2025 update to its terms of service and privacy policy, which prompted concerns about personal data collection and modding restrictions.\footnote{\url{https://www.gamesradar.com/games/borderlands/borderlands-games-are-still-being-review-bombed-on-steam-following-take-two-terms-of-service-backlash-but-modders-say-the-main-clause-everyones-freaking-out-about-was-already-there/}}
Although privacy concerns appear in all three communities, they overshadow gameplay in Chinese and Spanish reviews. This contrast suggests different sensitivities to data collection and publisher policies across language communities.

Interestingly, these concerns do not translate uniformly into overall Steam ratings. Italian and Spanish players remain relatively positive at 76.99 and 69.13, respectively, whereas Simplified Chinese players give an extremely low rating of 9.94. This suggests that Italian and Spanish players partly separate privacy concerns from their overall evaluation, while Chinese players more often treat them as decisive.

\subsubsection{Regional Treatment and Community Governance}

Regional access and community management can become central to game evaluations. In \textit{Dead by Daylight}, Simplified Chinese reviews criticize streaming bans and region-locked cosmetic collaborations, making regional restrictions the leading negative theme. In \textit{TEKKEN 7}, Chinese players criticize mainland China's exclusion from the Tekken World Tour and a Chinese competitor's disqualification, bringing esports management, company ethics, and company perception to the foreground. \textit{HELLDIVERS 2} shows a related pattern: Chinese players describe a large-scale war event as scripted and unwinnable despite collective effort, making developer relations and live operations dominant negative themes. Brazilian Portuguese reviews instead foreground gameplay and cooperative play. These cases suggest that cross-language differences extend to expectations of fair regional treatment and meaningful community participation.

\subsubsection{Divergent Responses to Commercial Practices}

Commercial practices carry different weights in game evaluations across language communities. In \textit{Call of Duty: Modern Warfare Remastered}, French reviews are more likely to frame controversial sales and monetization practices as an ethical betrayal, emphasizing corporate ethics and company attitude. Russian reviews instead focus more on remaster quality, visuals, and nostalgia, suggesting that these controversies carry less weight in their overall assessments.

\subsubsection{Political Tensions Manifested in Game Reception}

The Ukrainian-developed \textit{S.T.A.L.K.E.R.: Shadow of Chornobyl - Enhanced Edition} illustrates how geopolitical tensions enter game reception. Ukrainian players give the game a rating of 84.89 and express appreciation across almost all aspects, including graphics, localization, pricing, developers, and remaster quality. Russian players, by contrast, give a rating of 29.06 and predominantly criticize the same aspects. The main thematic difference is nostalgia in Ukrainian reviews versus explicit political references in Russian reviews, underscoring the geopolitical tensions surrounding this Ukrainian game.

\subsubsection{Games Achieving High Cross-Cultural Agreement}

Four titles stand out as rare cases of near-universal acclaim across languages, all with average Steam ratings above 97. \textit{Undertale} and \textit{DELTARUNE} receive particularly strong evaluations for music, art direction, emotional impact, and narrative, suggesting that their resonance travels well across cultures. By contrast, \textit{RimWorld} and \textit{Balatro} are praised for engagement, replayability, core gameplay, difficulty, and strategic depth, showing that carefully tuned, systems-driven play can attract consistently positive reception.
These cases suggest that emotionally resonant storytelling and engaging gameplay systems can both support positive reception across language communities.

\subsection{Implications for Cross-Cultural Game Research}

Taken together, these findings show that cross-language differences in game reception involve both how players evaluate an aspect and how much attention they give it. Similar sentiment can conceal different priorities, while overall Steam ratings can obscure disagreement about specific game aspects.

For researchers, our framework offers a scalable way to examine specific segments of the player base, focusing on particular games or language communities within a large review corpus. The shared codebook provides a consistent basis for comparing the same game aspects across languages, while theme frequencies and sentiment scores reveal differences in players' priorities and evaluations. Researchers can trace these patterns to initial codes and supporting quotes to examine how players express and explain their assessments. This connects large-scale quantitative comparisons with richer qualitative accounts of game reception, preserving the contextual evidence needed for human interpretation.
For developers, the framework could inform design before release through reviews of comparable games, and guide patches, localization improvements, and responses to player concerns after release.

%% file: sections/7_Limitation.tex
\section{Limitations}

\subsection{Representativeness of Data}

\paragraph{Platform and Reviewer Selection}
Our dataset is drawn exclusively from Steam and may not represent console or mobile players, whose expectations and evaluative norms may differ. Players who choose to post reviews may also differ from those who do not. Extending the framework to other platforms and complementing reviews with surveys or interviews could broaden its coverage.

\paragraph{Language as a Proxy for Cultural Communities}
We use review language as a practical proxy for cultural community, treating players who select the same language for their reviews as one group. However, language groups span nations, regions, and subcultures, limiting our ability to distinguish differences within a language. Our findings therefore apply to Steam language communities (e.g., ``Japanese-speaking players''), rather than all speakers of a language or entire national or ethnic groups.

\paragraph{Low-review Games}
Our framework requires enough reviews in multiple languages. As discussed in Section~\ref{sec:select}, games beyond roughly the top 2,000 best-selling Steam titles often have fewer than 10 reviews per non-English language. This biases our study toward popular, commercially successful games and limits its applicability to small studios or niche titles with few reviews. Analyzing these games may require proactive data collection across language communities to obtain sufficient evidence for cross-cultural comparison.

\subsection{LLM for Qualitative Research}

Consistent with prior work emphasizing researcher agency~\cite{jiang2021supporting,schroeder2025large}, our design uses LLMs to assist with coding and translation while retaining human responsibility for analytical guidance, decision-making, and interpretation. Our validation provides preliminary evidence of reliability within the evaluated samples, but does not cover all games and languages. LLM-generated codes and translations may also miss linguistic or cultural nuances. We therefore treat the codes as one structured interpretation that highlights cross-language patterns for human analysis, rather than an exhaustive account of meanings in each language.

\subsection{Analyzing Groups of Games}
Our framework currently compares language communities within individual games. An interesting extension would be to compare language communities across groups of games, such as those sharing a developer, publisher, or genre. Such comparisons could reveal whether divergences are genre-specific (e.g., competitive multiplayer versus narrative-driven indies) or reflect consistent developer or publisher practices.

%% file: sections/8_Conclusion.tex
\section{Conclusion}

We present the first study to combine large-scale coverage with explainable content analysis of cross-language game reviews, enabled by a human-AI collaborative framework. Our mixed-methods approach identifies games with potential cross-language differences and supports in-depth qualitative analysis, while retaining the flexibility to examine individual titles on demand.
Our findings show how narrative expectations, game mechanics and stability, localization quality, cultural proximity, and perceptions of developers and publishers shape cross-language reception. We also identify rare cases of broad agreement across languages.
We hope this work establishes a useful method for cross-language studies and informs future human-AI approaches to content analysis in HCI and social science.

%% file: sections/appendix.tex
\newpage
\appendix

\section{Quantitative Method for Selecting Games}\label{app:selection}

\subsection{Game-Language Rating Calculation}

To compare how different language communities evaluate the same game, we require a per-language rating that accounts for review reliability and avoids pooling biases.

For each \{game, language\} pair, we aggregate review votes to obtain a \textbf{Game-Language Rating}. For review $i$, we use the \emph{binary vote} $v_i \in \{0,100\}$ indicating a positive ($100$) or negative ($0$) recommendation, and we compute the rating as the mean of binary votes.

\paragraph{Game--language matrix.}
Organizing these ratings across games and languages, we form a \emph{game--language matrix} $R \in [0,100]^{G \times L}$. Rows index games $g$, columns index languages $\ell$, and each entry $R_{g\ell}$ is the Game--Language Rating. 

To ensure that each matrix entry reflects a stable estimate rather than noise from a handful of reviews, we treat $R_{g\ell}$ as \emph{missing} whenever game $g$ has fewer than 100 reviews in language $\ell$. We then apply a second filter at the game level to maintain comparability across languages: we retain only those games that have at least three languages meeting the 100-review threshold. Applying these two criteria to our initial set of 2,000 games yields 1,153 games, producing a $1153$ (games) $\times$ $30$ (languages) matrix with some missing cells. We refer to this matrix as the \textbf{sparse game--language matrix}. ``Sparse'' here denotes missing entries from insufficient reviews, not ratings of zero.

While the sparse matrix offers broad coverage, it can suffer from sample-size imbalance: games vary in the number of languages with sufficient reviews, and languages vary in the number of covered games. To control for this, we also construct a subset restricted to the ten languages with the largest total review counts. We keep games with more than 100 reviews in each of these languages (266 games in total), which we refer to as the \textbf{dense game--language matrix}.

\subsection{Overall Language Rating}\label{sec:overall}

We begin by summarizing per-language ratings across the initial game pool, before applying the review-count and language-coverage filters. For each language, we compute the mean and standard deviation of $R_{g\ell}$ across all games with reviews in that language.
Table~\ref{tab:avg_lang_rating} (left) reports the top and bottom five languages. Across this pool, the highest averages are observed for Romanian, Portuguese, Greek, Vietnamese, and Indonesian, whereas Turkish, Korean, Traditional Chinese, Japanese, and Simplified Chinese appear at the lower end. We also see a \emph{clustering} pattern: languages that are geographically or linguistically proximate (e.g., Romance languages) show similar mean levels, while East Asian languages cluster toward lower means. 

Because languages cover different sets of games in the initial pool, its averages conflate language-specific evaluation tendencies with differences in game coverage. To provide a more balanced view, we compute per-language averages on the dense game--language matrix, which includes the same 266 games for each of the top 10 languages. As shown in Table~\ref{tab:avg_lang_rating} (right), the ranking compresses (smaller SDs approximately 9 to 18) and reorders: Brazilian Portuguese and Spanish appear at the top; English, French, German, Polish, Russian, and Turkish form a narrow mid-80s band; while Korean and Simplified Chinese remain the lowest even under the controlled game set. The upward shift for Turkish relative to the initial pool indicates that part of the cross-language gap is attributable to differences in game coverage rather than evaluation alone.

\begin{table*}[t]
\centering

\begin{minipage}{0.48\textwidth}
\centering
\begin{tabular}{lcccc}
\hline
Language & Mean $\downarrow$ & SD & \#Games & \#Reviews \\
\hline
Romanian   & 92.37 & 18.40 & 1,038 &   14,826 \\
Portuguese & 91.98 & 16.64 & 1,552 &   41,636 \\
Greek      & 91.82 & 19.12 & 1,121 &    9,161 \\
Vietnamese & 91.31 & 20.16 &  945 &   11,909 \\
Indonesian & 90.86 & 21.85 &  841 &    8,986 \\
\hline
Turkish    & 83.38 & 19.04 & 1,792 &  402,537 \\
Korean    & 81.41 & 19.63 & 1,804 &  288,055 \\
Chinese (T)   & 78.55 & 23.05 & 1,757 &  150,976 \\
Japanese   & 77.09 & 23.18 & 1,706 &   99,948 \\
Chinese (S)   & 76.38 & 20.44 & 1,927 & 2,682,878 \\
\hline
\end{tabular}
\end{minipage}%
\hfill
\begin{minipage}{0.48\textwidth}
\centering
\begin{tabular}{lcccc}
\hline
Language & Mean $\downarrow$ & SD & \#Games & \#Reviews \\
\hline
Brazilian & 91.59 &  9.50 & 266 &  561,587 \\
Spanish   & 90.81 & 10.44 & 266 &  515,966 \\
Polish    & 88.72 & 11.30 & 266 &  249,411 \\
French    & 87.94 & 11.62 & 266 &  265,813 \\
Turkish   & 86.58 & 10.59 & 266 &  307,279 \\
German    & 86.46 & 12.32 & 266 &  339,723 \\
English   & 86.16 & 12.75 & 266 & 4,974,378 \\
Russian   & 86.03 & 12.03 & 266 & 1,694,381 \\
Korean   & 82.20 & 15.37 & 266 &  190,817 \\
Chinese (S)  & 79.08 & 17.65 & 266 & 1,950,029 \\

\hline
\end{tabular}
\end{minipage}
\caption{Summary statistics of game ratings by language. \textbf{Left:} Top 5 and bottom 5 languages by mean rating across the initial game pool. \textbf{Right:} Results for all 10 languages from the dense game--language matrix. ``Brazilian'' denotes Brazilian Portuguese. ``Chinese (S)'' and ``Chinese (T)'' denote Simplified and Traditional Chinese, respectively.
}\label{tab:avg_lang_rating}
\end{table*}

\subsection{Metrics}

\subsubsection{Cross-Language Rating Variation}

We quantify variation in each game’s ratings across languages to identify cases of high disagreement or agreement. Let $R_{g\ell} \in [0,100]$ denote the Game–Language Rating for game $g$ in language $\ell$. Let $L_g$ be the set of languages with sufficient reviews for game $g$ in the matrix under consideration (sparse or dense), and let
\[
\mu_g \;=\; \frac{1}{|L_g|}\sum_{\ell \in L_g} R_{g\ell}
\]
be the mean rating of game $g$ across its observed languages. For each game $g$ we compute three metrics:

\begin{itemize}
  \item \textbf{Range gap}:
  \[
  \mathrm{RG}_g \;=\; \max_{\ell \in L_g} R_{g\ell} \;-\; \min_{\ell \in L_g} R_{g\ell}.
  \]
  \item \textbf{Standard deviation}:
  \[
  \mathrm{SD}_g \;=\; \sqrt{\frac{1}{|L_g|-1}\sum_{\ell \in L_g}\left(R_{g\ell}-\mu_g\right)^2 }.
  \]
  \item \textbf{Coefficient of variation}:
  \[
  \mathrm{CV}_g \;=\; \frac{\mathrm{SD}_g}{\mu_g}.
  \]
\end{itemize}

We compute these metrics on both the sparse and dense game–language matrices and use them to identify games that exhibit potential cross-language abnormalities warranting in-depth examination. Because $|L_g|$ varies in the sparse matrix, the metrics have known sensitivities: the range gap increases with language count due to a higher chance of extremes; the standard deviation shows a moderate upward bias with broader language coverage; the coefficient of variation normalizes by the mean but can overemphasize low-mean games. Our goal, however, is to surface candidates with clear and interpretable cross-language patterns rather than to produce a bias-free league table. These sensitivities do not undermine the usefulness of the sparse matrix for identifying games that merit closer analysis.

\subsubsection{Ranking Outliers}

We next examine outliers along the language axis: games that rank unusually high or low within a given language. To enable direct comparison, we use the dense game--language matrix, which contains the same 266 games for each of the top 10 languages. The procedure is:

\begin{enumerate}
  \item For each language $\ell$, rank the 266 games by $R_{g\ell}$ from highest to lowest to obtain ranks $r_{g\ell} \in \{1,\dots,266\}$.
  \item Compute a language-agnostic baseline rank for each game,
  \[
  \bar{r}_g \;=\; \frac{1}{10}\sum_{\ell} r_{g\ell}.
  \]
  \item Define the \textbf{rank difference} for each pair $(g,\ell)$ as
  \begin{equation}\label{eq:rank}
        \Delta r_{g\ell} \;=\; \bar{r}_g \;-\; r_{g\ell}.
  \end{equation}

\end{enumerate}
A positive $\Delta r_{g\ell}$ indicates that game $g$ ranks higher in language $\ell$ than its overall baseline (favored in that language). A negative value indicates the game ranks lower than its baseline (disfavored in that language).

Using the same game set across languages removes confounds from differing game coverage. Ranking within each language makes the analysis invariant to scale differences in rating levels and less sensitive to the magnitude of extreme ratings. Using the cross-language mean rank as a baseline helps account for shared assessments of a game's intrinsic quality, highlighting language-specific deviations from its overall standing. This provides a comparable signal across languages for selecting cases for deeper analysis.

\subsection{Results}

\subsubsection{Cross-Language Disagreement and Agreement}

\paragraph{Highest Disagreement}

In the sparse game--language matrix, biases arising from varying sample sizes led us to manually inspect the language distribution and review patterns of the top 5 games under each of the three disagreement metrics: range gap, standard deviation, and coefficient of variation. After excluding games and languages with insufficient samples (e.g., languages represented by about 100 short reviews), we retained 7 games along with the languages having the highest and lowest ratings for each game.
In the dense game--language matrix, the three metrics were highly correlated, with all pairwise Pearson correlations exceeding \(r = 0.96\) (\(p < 0.001\)). Consequently, we considered only the top 5 games with the largest range gaps to identify cases of high disagreement for further analysis.

\paragraph{Highest Agreement}

The games with the lowest disagreement metrics in the sparse game--language matrix were overly biased toward games represented in only 3 languages and by very few reviews. We therefore use only the dense game--language matrix to identify games with the highest cross-language agreement. Surprisingly, all three cross-language disagreement metrics identified the same five games as having the lowest disagreement, all with average ratings above 97 and range gaps below 2 across the top 10 languages. These games are Wallpaper Engine, RimWorld, Balatro, Undertale, and DELTARUNE. Excluding Wallpaper Engine, which is more of a creative utility than a game, the remaining titles with the highest cross-language agreement are all indie single-player games.

The first and second sections of Table~\ref{tab:games_languages} present the selected games with the highest and lowest levels of cross-language disagreement.

\subsubsection{Ranking Outliers}

Using the rank differences defined in Eq.~\ref{eq:rank}, we select game--language pairs satisfying both criteria: (1) within each language, the pair $(g,\ell)$ has the largest positive or largest negative $\Delta r_{g\ell}$, and (2) its absolute rank difference exceeds 90. These criteria identify pronounced deviations for closer qualitative analysis.

The third section of Table~\ref{tab:games_languages} summarizes the selected outliers. Positive cases show games that are markedly more appreciated within a specific language community relative to their overall standing; negative cases show the converse.

\section{LLM Prompt Templates}\label{app:prompt}

Here, we provide the prompt templates used for all LLM calls in our workflow.
Table~\ref{tab:prompt1} presents the prompt for generating initial codes from individual reviews.
Table~\ref{tab:prompt2} presents the prompt for assigning initial codes to existing themes.
Table~\ref{tab:prompt3} presents the prompt for proposing new themes from uncategorized initial codes.

\begin{table*}[t]
\centering
\begin{tabular}{|p{0.95\textwidth}|}
\hline
\textbf{System prompt:} \\
\hline
You are an experienced qualitative researcher conducting inductive coding of multilingual game reviews. Identify the aspects of game assessment expressed in the input review and represent them as concise, specific, non-overlapping initial codes in English. Support each code with evidence from the review, an English explanation, and an aspect-specific sentiment score. Do not introduce information unsupported by the review. \\
\hline
\textbf{User prompt:} \\
\hline
\#\# TASK

Read the review and identify all salient aspects of how the reviewer assesses the game. These may include mechanics, systems, difficulty, interface and user experience, performance, audio, art, narrative, progression, multiplayer, controls, accessibility, bugs, monetization, live operations, community, characters, balance, social features, and ethics. This list is illustrative, not exhaustive. When proposing new codes, use a level of granularity similar to the aspect examples listed above. \\ 

\#\# CODING RULES

- Each code must identify one specific aspect of game assessment supported by the review.\\
- Write code names in English, preferably using fewer than four words.\\
- Use neutral wording for code names. Represent the reviewer's attitude through the sentiment score.\\
- Avoid overlapping codes. Merge synonymous or near-duplicate codes while retaining distinct aspects.\\
- Code only content relevant to game assessment; not every sentence requires a code.\\
- If the review contains no sufficiently clear aspect to code, return an empty ``codes\_object'' array.\\

\#\# EVIDENCE AND EXPLANATION

For each code:

- Select one to three verbatim excerpts supporting the code. Preserve the original wording, punctuation, and emojis.\\
- Translate each non-English excerpt into English. If an excerpt is in English, set its translation field to null.\\
- Provide a short English explanation that clarifies the aspect and explains how the evidence supports the code.\\
- Ensure that the code, explanation, evidence, and sentiment score refer to the same aspect.\\

\#\# SENTIMENT

Assign one integer score to each code based on the attitude expressed toward that aspect: 1 = very negative, 2 = negative, 3 = neutral or mixed, 4 = positive, and 5 = very positive.

Interpret the evidence in the context of the review. If there are no clear sentiment cues, assign 3.

\#\# OUTPUT FORMAT

Return only valid JSON in the following structure, with no additional text or fields. Replace the placeholders with the corresponding values.

\quad ``codes\_object'': [\\
\quad\quad \{\\
\quad\quad\quad ``quotes'': [\\
\quad\quad\quad\quad \{\\
\quad\quad\quad\quad\quad ``quote'': ``\textless verbatim excerpt from the review\textgreater'',\\
\quad\quad\quad\quad\quad ``translation'': \textless English translation string or null\textgreater\\
\quad\quad\quad\quad \},\\
\quad\quad\quad\quad \textless additional quote objects, if any\textgreater\\
\quad\quad\quad ],\\
\quad\quad\quad ``explanation'': ``\textless one- to two-sentence English explanation\textgreater'',\\
\quad\quad\quad ``code'': ``\textless concise English code name\textgreater'',\\
\quad\quad\quad ``score'': \textless integer sentiment score from 1 to 5\textgreater\\
\quad\quad \},\\
\quad\quad \textless additional code objects, if any\textgreater\\
\quad ]\\

\#\# INPUT REVIEW

\textless Insert a single game review in any language.\textgreater\\
\hline
\end{tabular}
\caption{Prompt for generating initial codes from individual reviews.}\label{tab:prompt1}
\end{table*}

\begin{table*}[t]
\centering
\begin{tabular}{|p{0.95\textwidth}|}
\hline
\textbf{System prompt:} \\
\hline
You are an experienced qualitative researcher assigning initial codes from game reviews to an established theme-level codebook. For each input code, select exactly one existing theme when there is a clear match. Otherwise, assign ``uncategorized''. Treat the existing themes as fixed and base each decision on the code's meaning and supporting evidence.\\
\hline
\textbf{User prompt:} \\
\hline
\#\# TASK

Assign each input initial code to one existing theme or to ``uncategorized''. \\

\#\# INPUT DESCRIPTION

The input contains:

- ``themes'': The current theme-level codebook. Each theme has a name and example code names previously assigned to it.\\
- ``initial\_codes'': A list of unassigned initial codes. Each entry contains a unique code name, representative evidence quotes, and their corresponding explanations. \\

\#\# ASSIGNMENT PROCEDURE

For each initial code:

1. Read its name, evidence quotes, and explanations to determine its meaning in context.\\
2. Compare that meaning with the theme names and example code names in the existing codebook.\\
3. If one theme is clearly the best fit, assign the code to that theme. If no clear match exists, assign ``uncategorized''.\\
4. Provide a brief rationale for the assignment. \\

\#\# RULES

- Base assignments on meaning in context, rather than code-name similarity alone.\\
- Treat the existing themes as fixed: do not create, rename, merge, or split them.\\
- Return exactly one assignment for every input code name. Do not omit codes or introduce additional codes.\\
- Copy code names and assigned theme names exactly as they appear in the input.\\
- Multiple codes may belong to the same theme, but each code must receive only one assignment.\\
- Be conservative when in doubt: prefer ``uncategorized'' over a weak or speculative match.\\

\#\# OUTPUT FORMAT

Return only a valid JSON array in the following structure, with no additional text or fields. Include one object per input code and replace the placeholders with the corresponding values.\\

[\\
\quad \{\\
\quad\quad ``code'': ``\textless exact input code name\textgreater'',\\
\quad\quad ``rationale'': ``\textless brief explanation of the assignment\textgreater'',\\
\quad\quad ``theme'': ``\textless exact existing theme name or uncategorized\textgreater''\\
\quad \},\\
\quad \textless additional assignment objects, if any\textgreater\\
]\\

\#\# INPUT

\textless current themes and list of unassigned initial codes\textgreater\\
\hline

\end{tabular}
\caption{Prompt for assigning initial codes to existing themes.}\label{tab:prompt2}
\end{table*}

\begin{table*}[t]
\centering
\begin{tabular}{|p{0.95\textwidth}|}
\hline
\textbf{System prompt:} \\
\hline
You are an experienced qualitative researcher developing a theme-level codebook for game reviews. Group uncategorized initial codes into new themes using the supplied codebook as a reference. \\
\hline

\textbf{User prompt:} \\
\hline
\#\# TASK \\

Propose new themes by grouping the uncategorized initial codes. Use the existing codebook to guide the level of abstraction and naming of new themes. \\
This step does not require every uncategorized code to be assigned. Do not force codes into unsuitable themes to achieve complete coverage. \\

\#\# INPUT DESCRIPTION \\

The input contains:

- Uncategorized initial codes, each with a unique code name, representative evidence quotes, and their corresponding explanations.\\
- The existing codebook, containing each theme name and example code names previously assigned to it. \\

\#\# MERGING PRINCIPLES \\

- Semantic Equivalence: Merge codes only when they have similar meanings within the gaming context. For example, ``soundtrack'' and ``background music'' may be grouped under ``Music''. Related but distinct aspects, such as ``Voice Acting'', must remain separate.\\
- Mutual Exclusivity: New themes must not overlap with existing themes or with one another.\\
- Contextual Fidelity: Each proposed theme must accurately reflect the aspects described in the original evidence quotes. Use the corresponding explanations to interpret the evidence rather than relying on code names alone. \\

\#\# RULES \\

- Give each new theme a concise English name that accurately represents its included codes.\\
- Copy code names exactly from the uncategorized input. Do not introduce additional codes.\\
- Place each included code under only one proposed theme. \\

\#\# OUTPUT FORMAT \\

Return only newly proposed themes as a valid JSON array in the following structure, with no additional text or fields. Include one object per new theme and replace the placeholders with the corresponding values.\\

[\\
\quad \{\\
\quad\quad ``theme'': ``\textless new theme name\textgreater'',\\
\quad\quad ``codes'': [\\
\quad\quad\quad ``\textless exact initial code name\textgreater'',\\
\quad\quad\quad \textless additional code names, if any\textgreater\\
\quad\quad ]\\
\quad \},\\
\quad \textless additional new theme objects, if any\textgreater\\
]\\

\#\# UNCATEGORIZED INITIAL CODES

\textless uncategorized codes with supporting quotes and explanations\textgreater\\

\#\# EXISTING CODEBOOK

\textless existing theme names and their example code names\textgreater\\

\hline
\end{tabular}
\caption{Prompt for proposing new themes from uncategorized initial codes.}\label{tab:prompt3}
\end{table*}

\clearpage
\onecolumn
\section{Visualization of Results for All Games of Interest}\label{app:results}

\begin{center}
\begin{minipage}{\linewidth}
  \centering
  \includegraphics[width=0.8\linewidth]{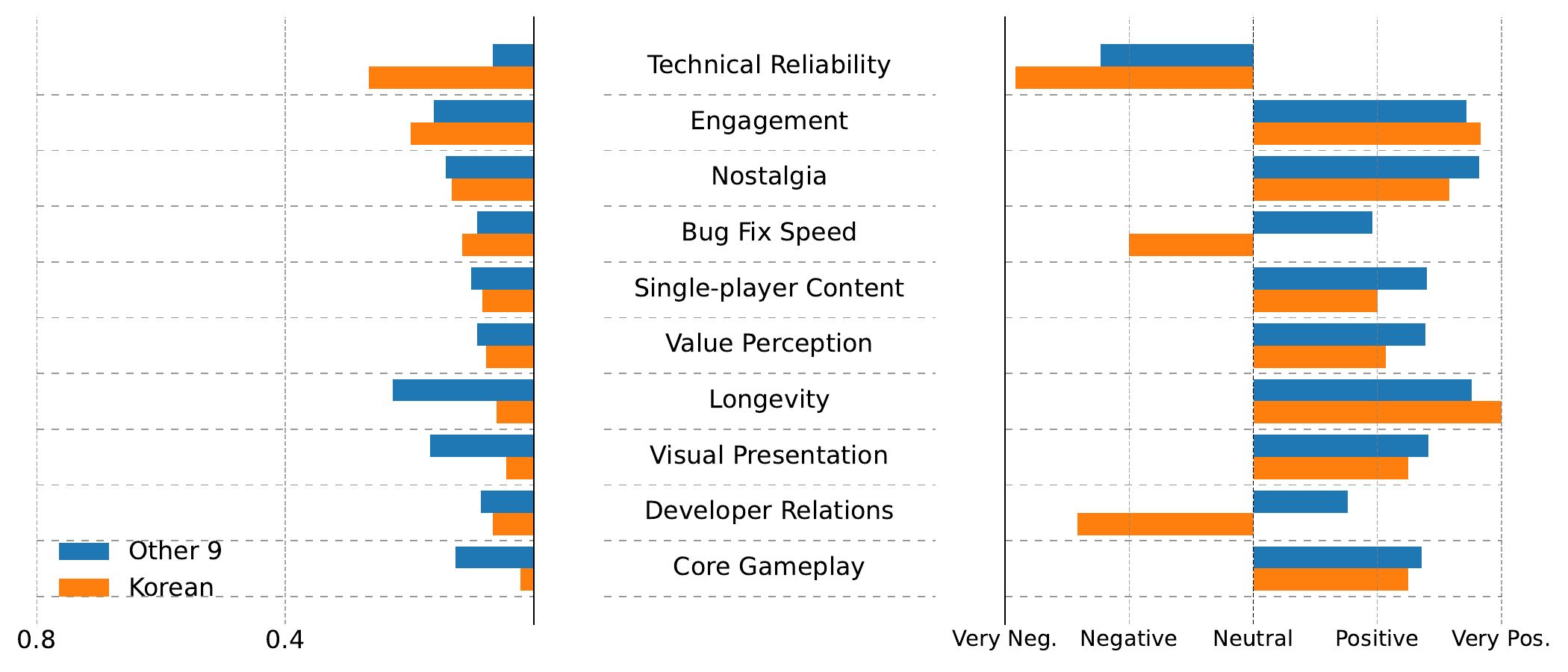}
  \captionof{figure}{Age of Empires II}
  \label{fig:AgeofEmpiresII}
\end{minipage}
\end{center}

\begin{center}
\begin{minipage}{\linewidth}
  \centering
  \includegraphics[width=0.8\linewidth]{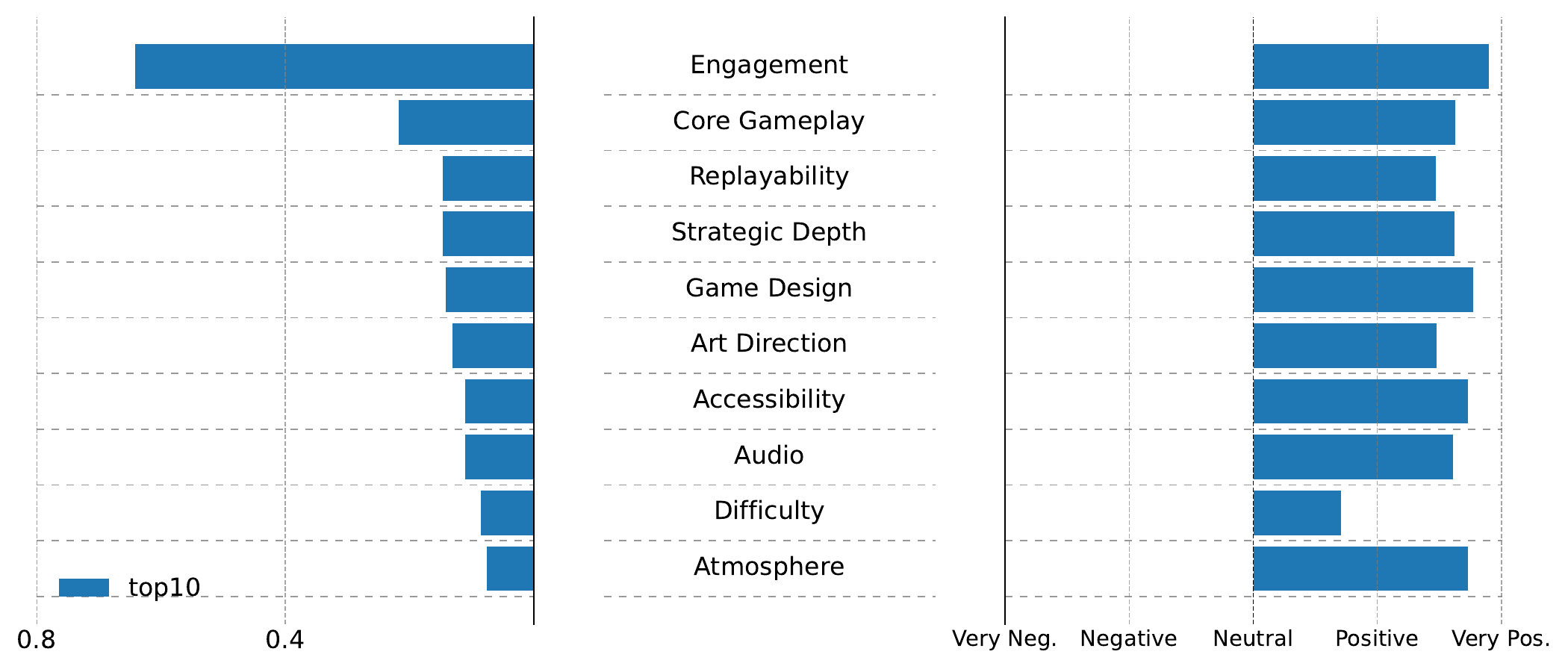}
  \captionof{figure}{Balatro}
  \label{fig:Balatro}
\end{minipage}
\end{center}

\begin{center}
\begin{minipage}{\linewidth}
  \centering
  \includegraphics[width=0.8\linewidth]{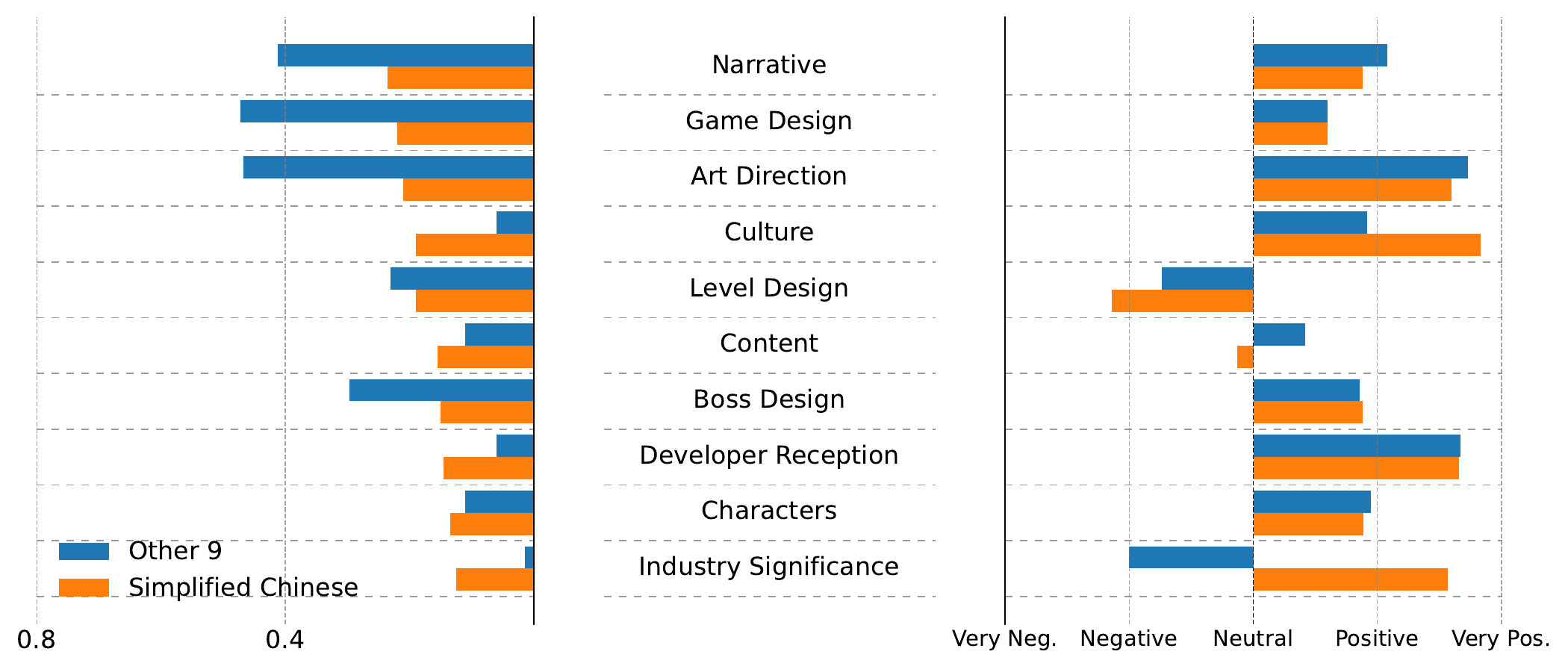}
  \captionof{figure}{Black Myth: Wukong}
  \label{fig:BlackMythWukong}
\end{minipage}
\end{center}

\begin{center}
\begin{minipage}{\linewidth}
  \centering
  \includegraphics[width=0.8\linewidth]{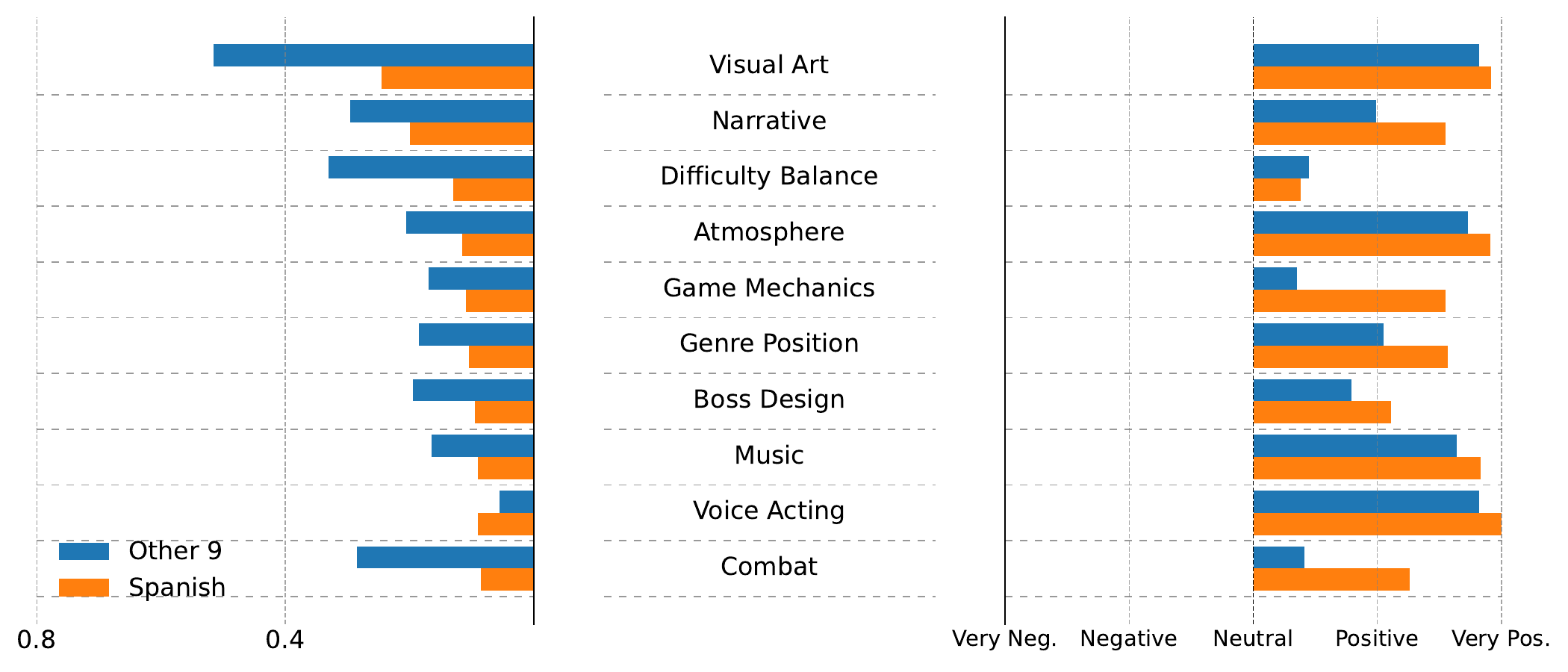}
  \captionof{figure}{Blasphemous}
  \label{fig:Blasphemous}
\end{minipage}
\end{center}

\begin{center}
\begin{minipage}{\linewidth}
  \centering
  \includegraphics[width=0.8\linewidth]{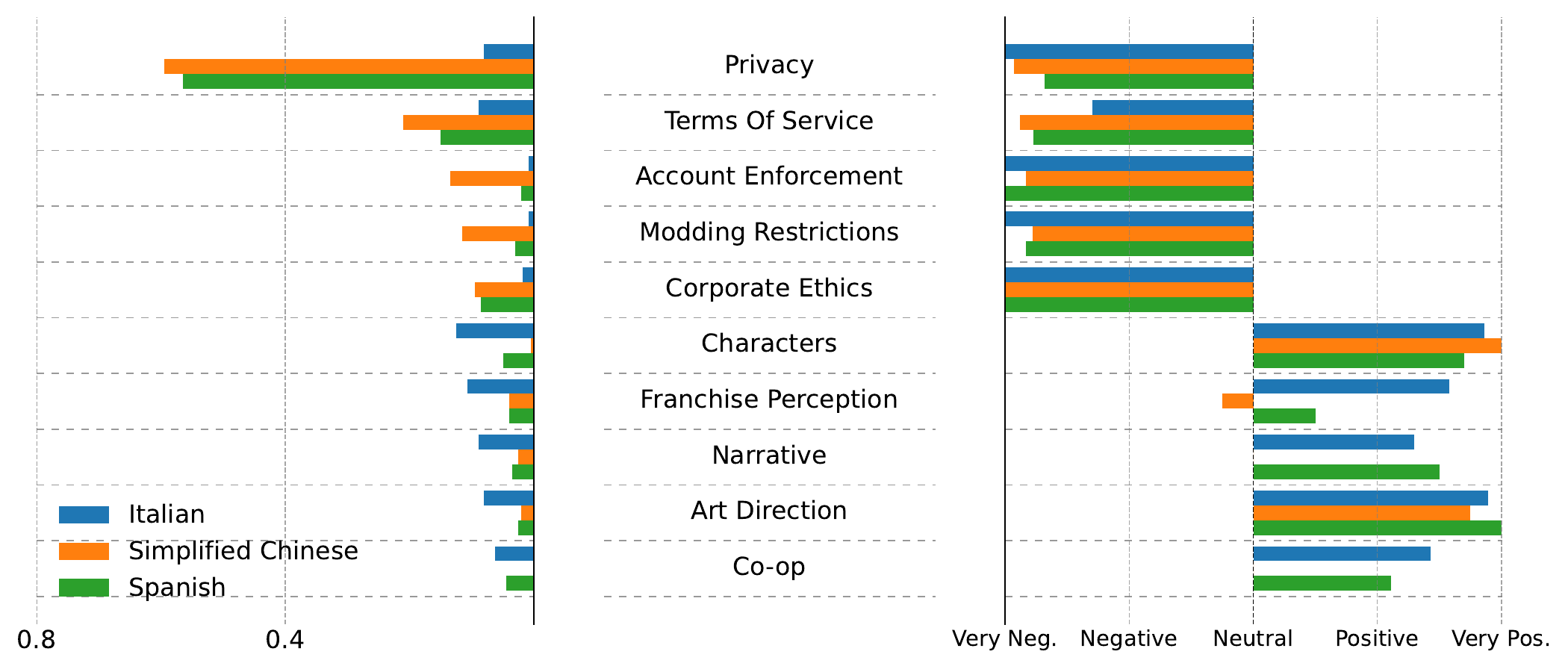}
  \captionof{figure}{Borderlands 2}
  \label{fig:borderlands2}
\end{minipage}
\end{center}

\begin{center}
\begin{minipage}{\linewidth}
  \centering
  \includegraphics[width=0.8\linewidth]{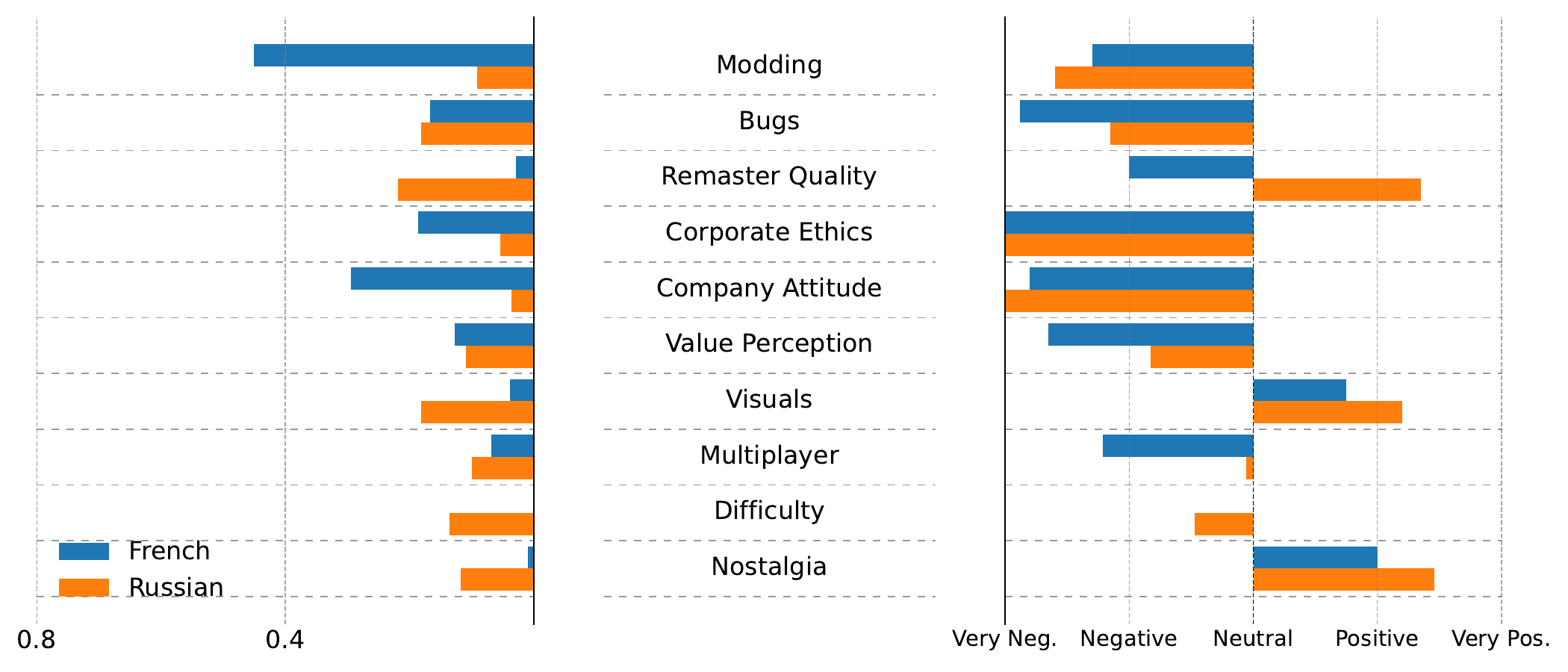}
  \captionof{figure}{Call of Duty: Modern Warfare}
  \label{fig:CallofDutyModernWarfare}
\end{minipage}
\end{center}

\begin{center}
\begin{minipage}{\linewidth}
  \centering
  \includegraphics[width=0.8\linewidth]{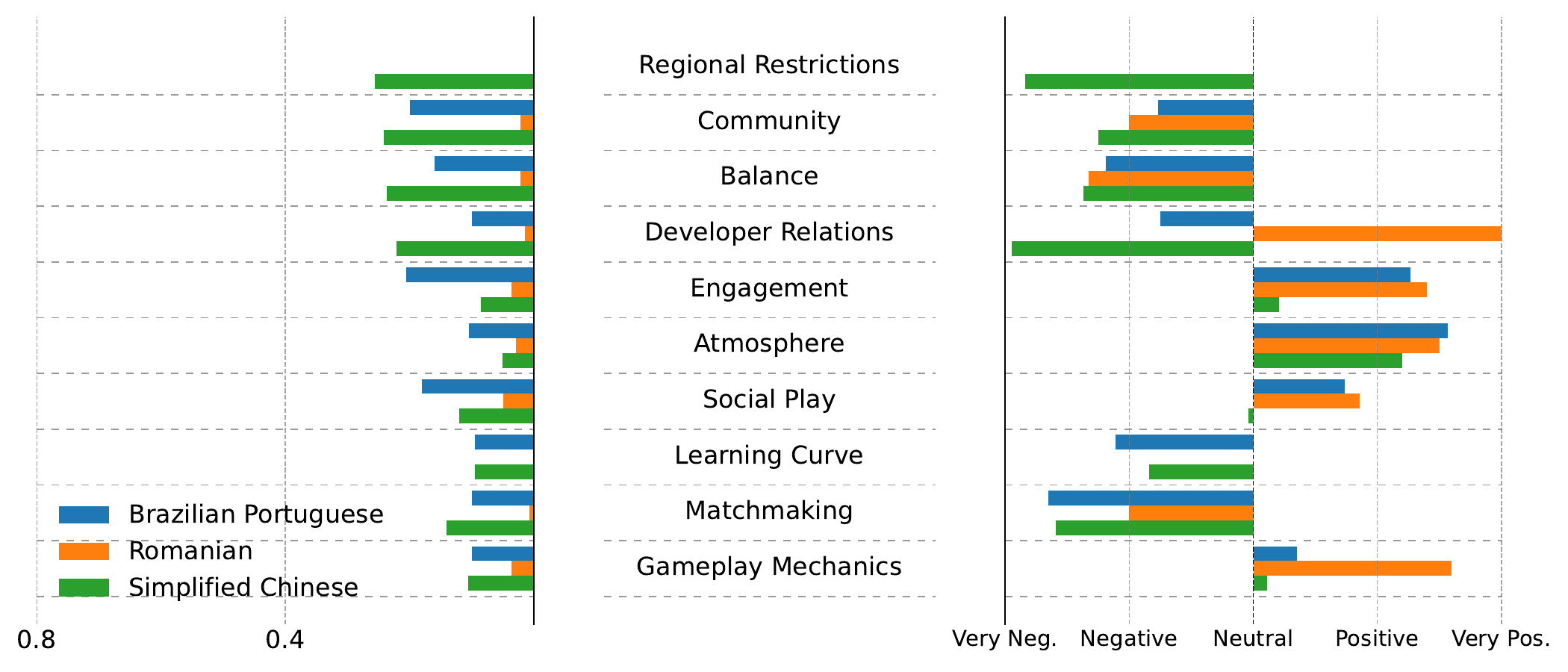}
  \captionof{figure}{Dead by Daylight}
  \label{fig:deadbydaylight}
\end{minipage}
\end{center}

\begin{center}
\begin{minipage}{\linewidth}
  \centering
  \includegraphics[width=0.8\linewidth]{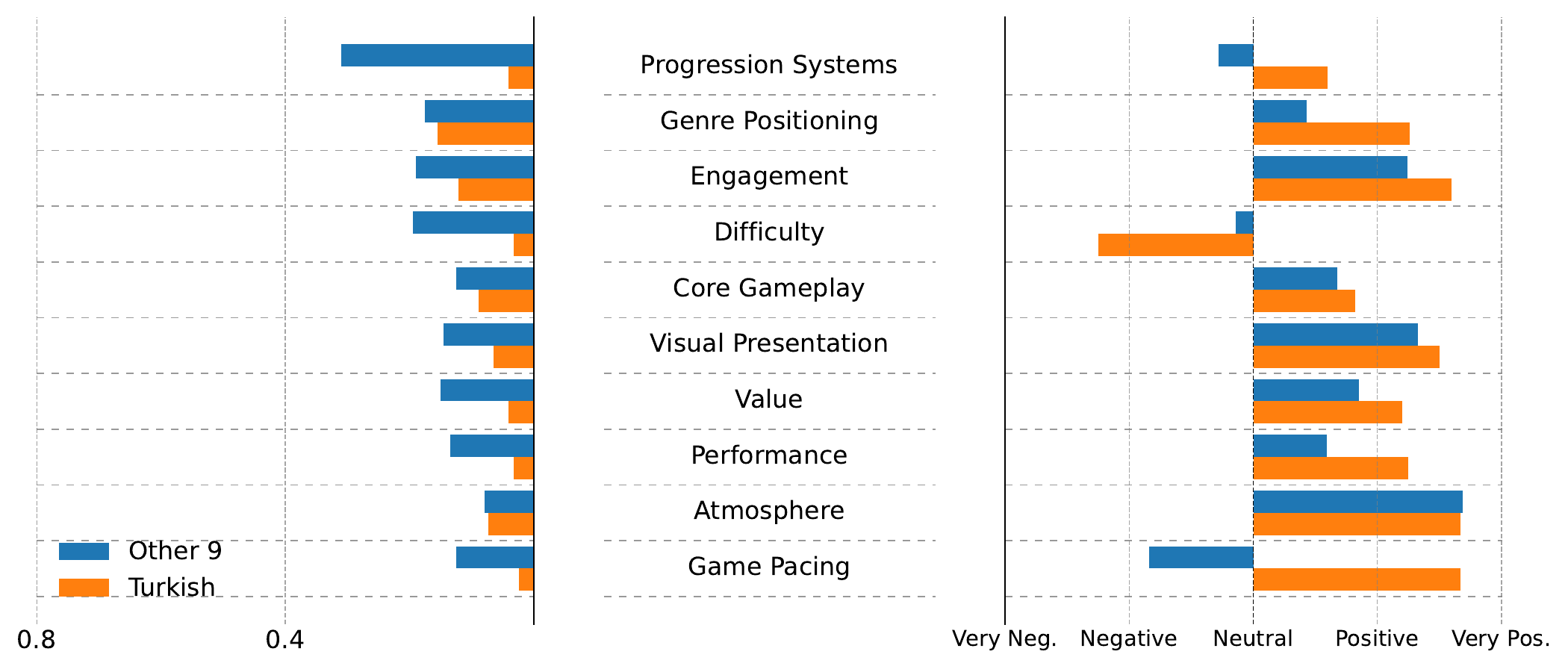}
  \captionof{figure}{Deep Rock Galactic: Survivor}
  \label{fig:DeepRockGalactic}
\end{minipage}
\end{center}

\begin{center}
\begin{minipage}{\linewidth}
  \centering
  \includegraphics[width=0.8\linewidth]{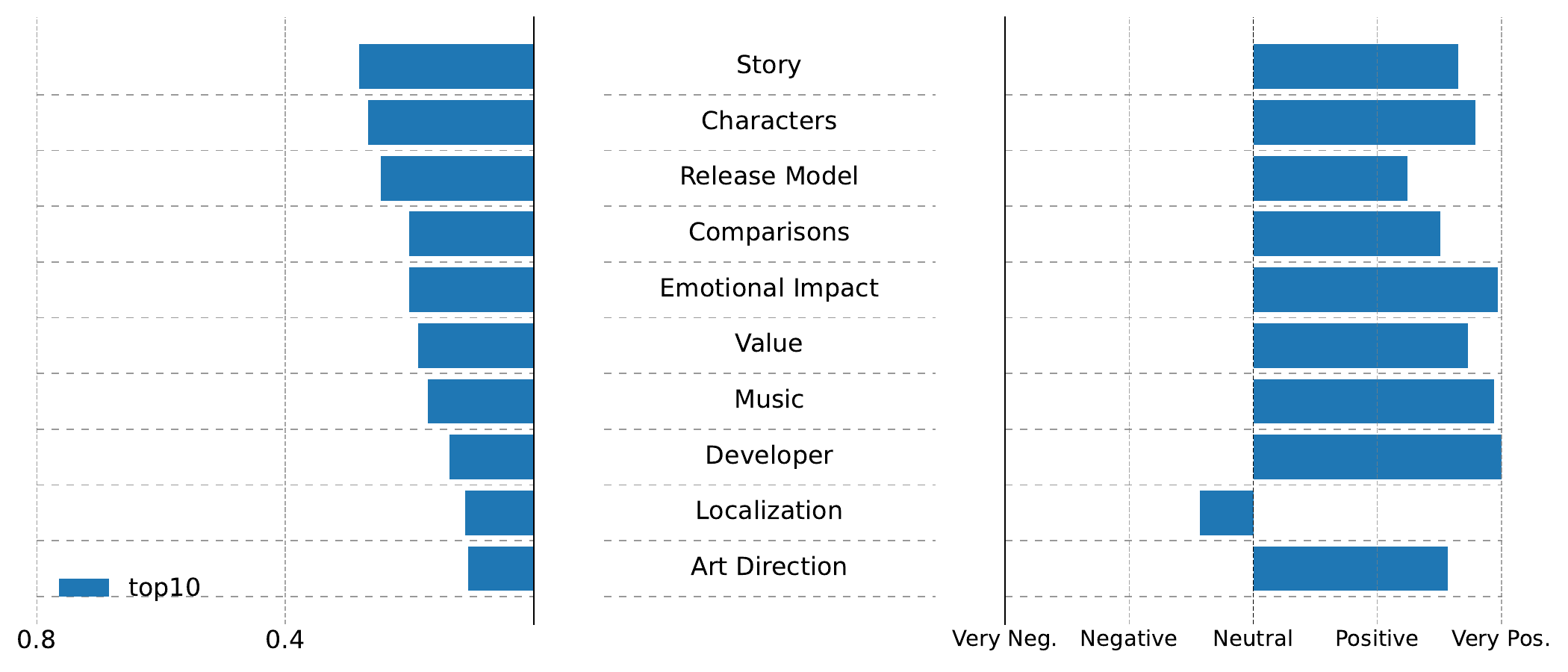}
  \captionof{figure}{DELTARUNE}
  \label{fig:Deltarune}
\end{minipage}
\end{center}

\begin{center}
\begin{minipage}{\linewidth}
  \centering
  \includegraphics[width=0.8\linewidth]{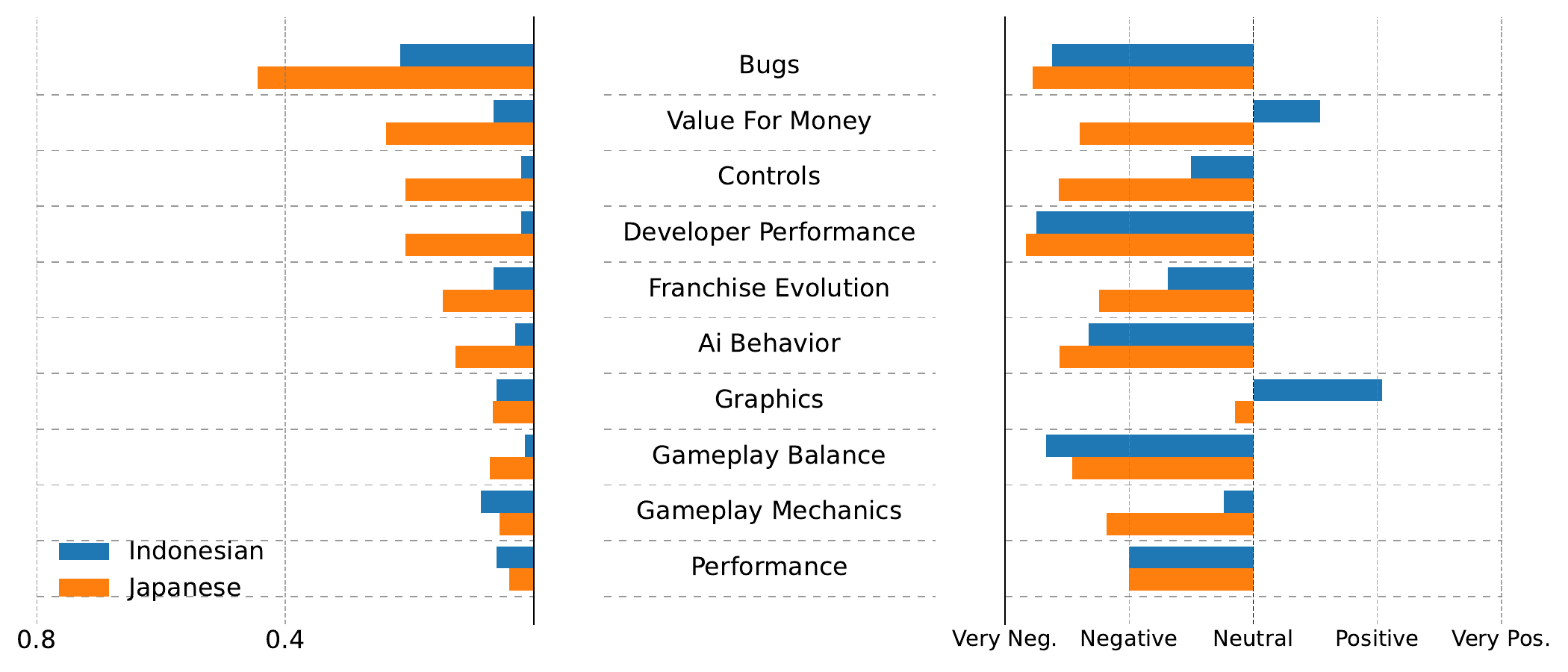}
  \captionof{figure}{EA SPORTS FC 25}
  \label{fig:EAFC25}
\end{minipage}
\end{center}

\begin{center}
\begin{minipage}{\linewidth}
  \centering
  \includegraphics[width=0.8\linewidth]{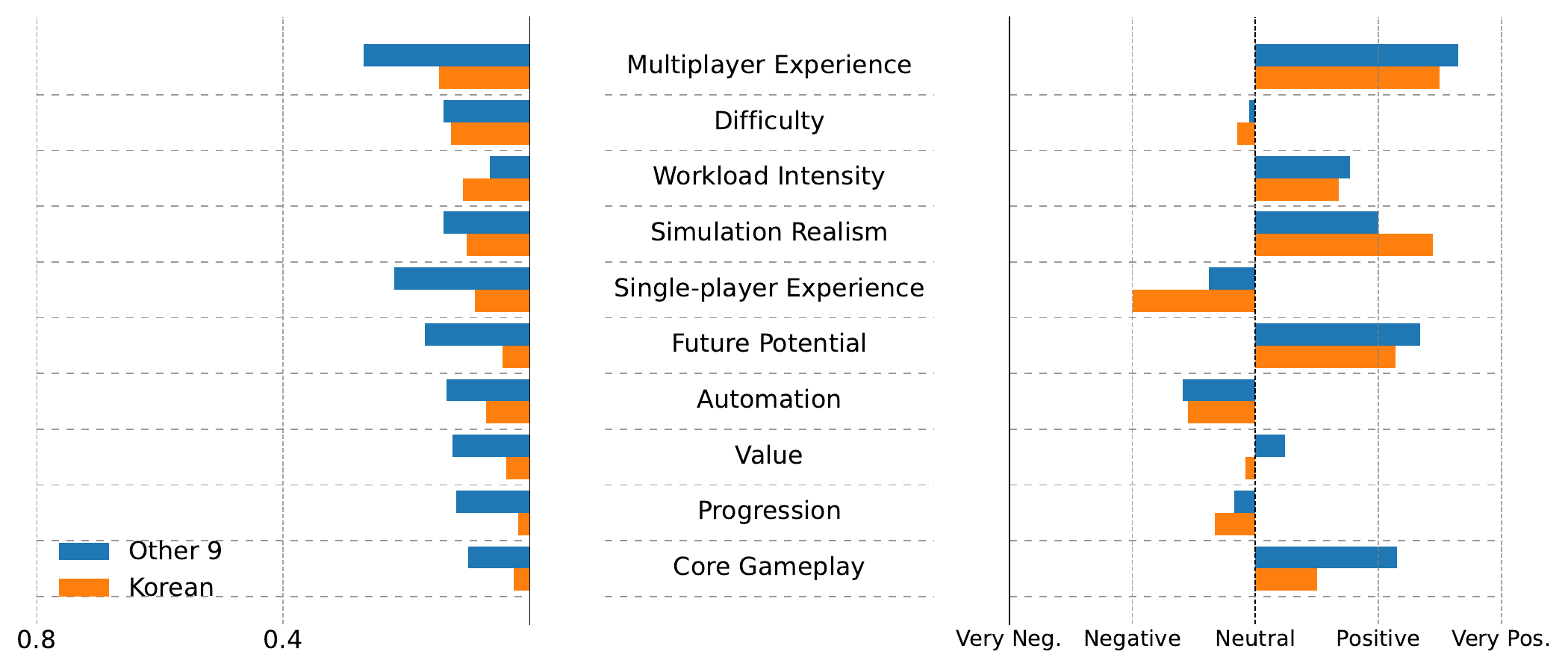}
  \captionof{figure}{Fast Food Simulator}
  \label{fig:fastfoodsimulator}
\end{minipage}
\end{center}

\begin{center}
\begin{minipage}{\linewidth}
  \centering
  \includegraphics[width=0.8\linewidth]{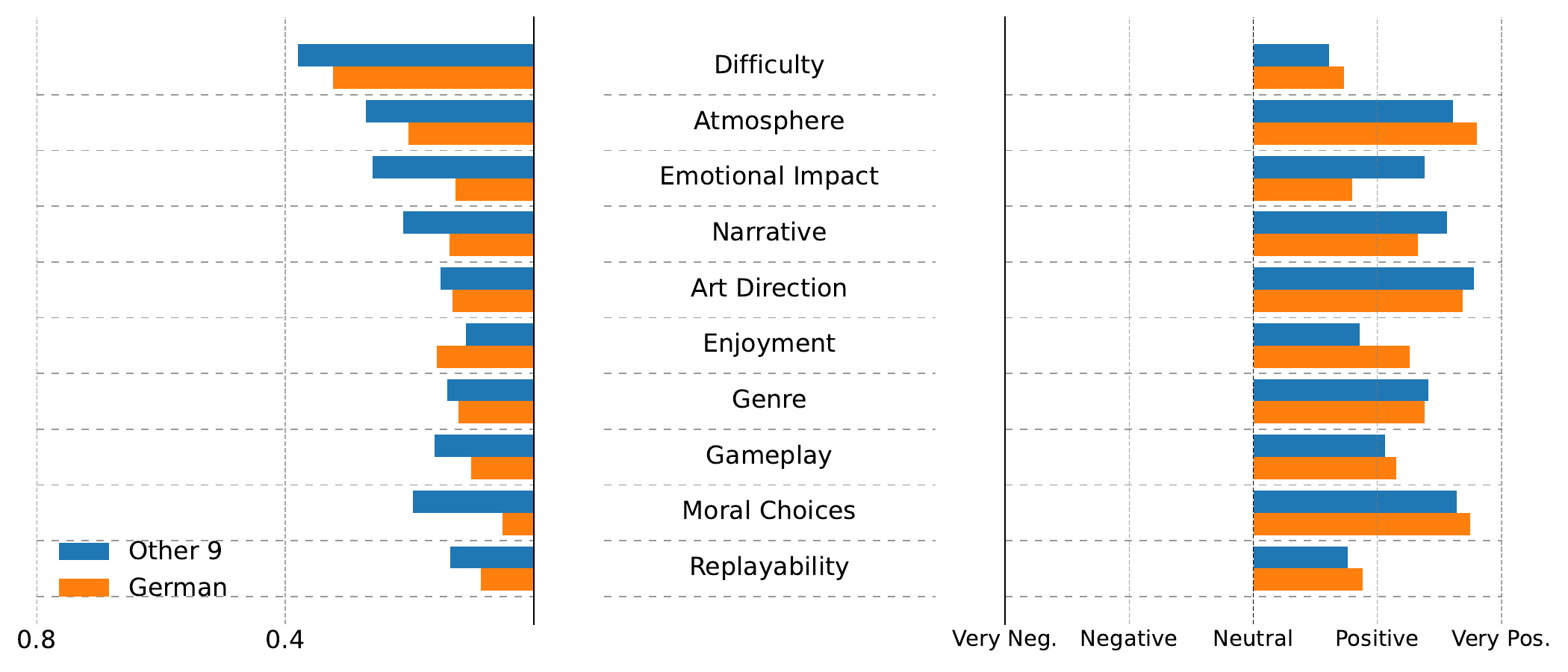}
  \captionof{figure}{Frostpunk}
  \label{fig:frostpunk}
\end{minipage}
\end{center}

\begin{center}
\begin{minipage}{\linewidth}
  \centering
  \includegraphics[width=0.8\linewidth]{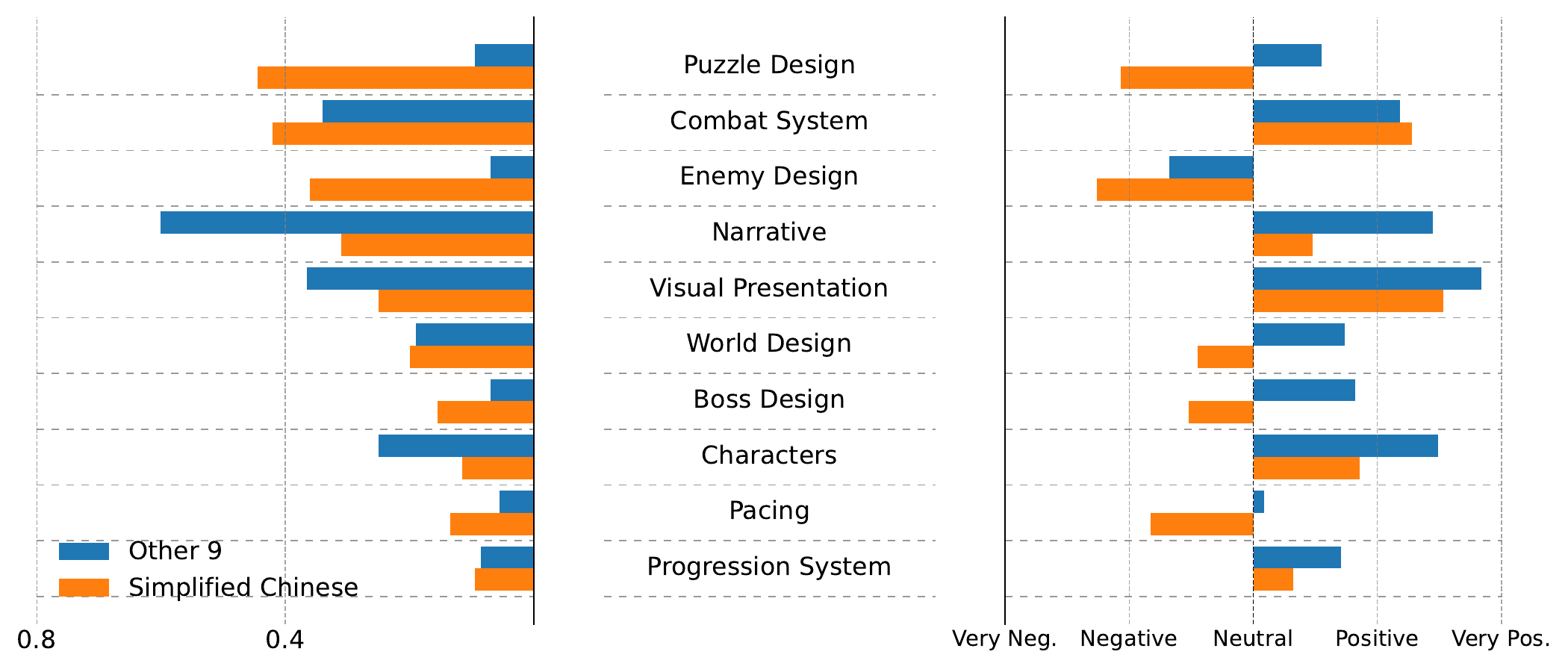}
  \captionof{figure}{God of War}
  \label{fig:GodofWar}
\end{minipage}
\end{center}

\begin{center}
\begin{minipage}{\linewidth}
  \centering
  \includegraphics[width=0.8\linewidth]{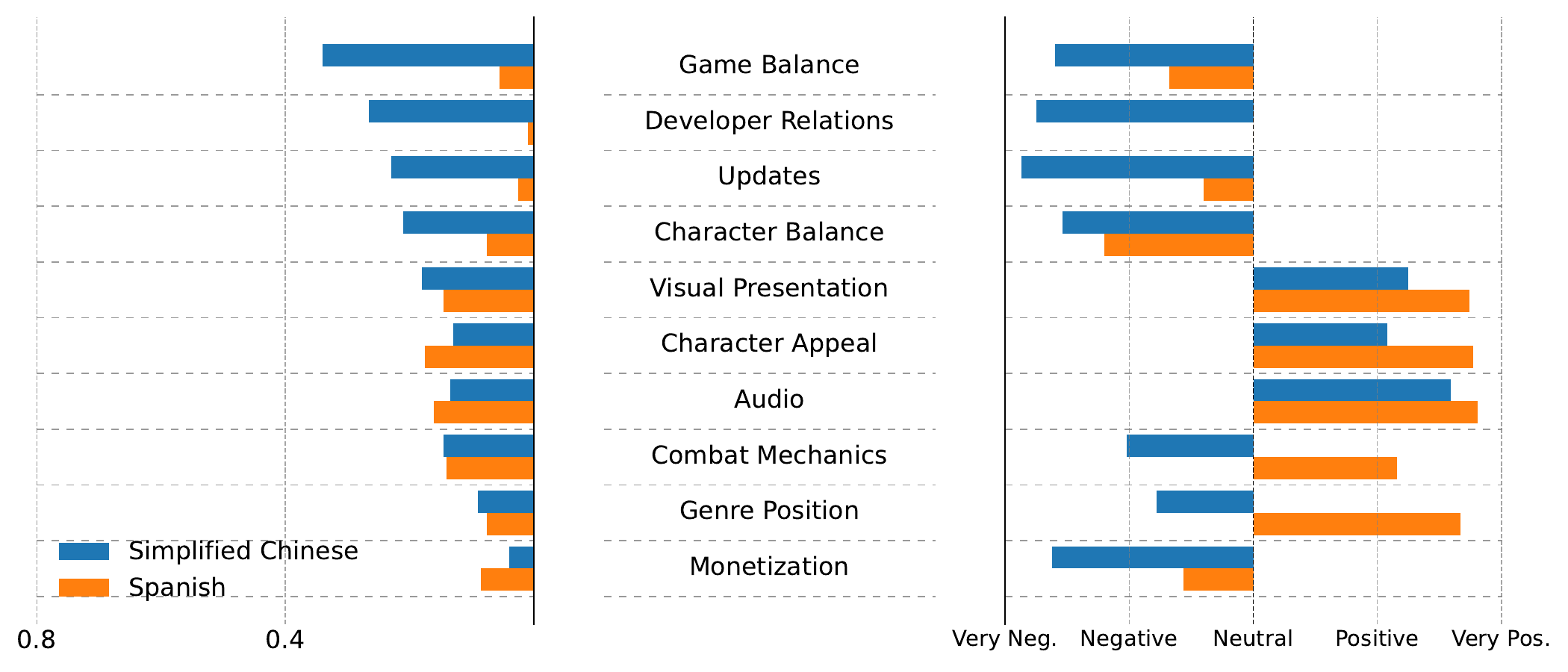}
  \captionof{figure}{GUILTY GEAR -STRIVE-}
  \label{fig:GuiltyGearStrive}
\end{minipage}
\end{center}

\begin{center}
\begin{minipage}{\linewidth}
  \centering
  \includegraphics[width=0.8\linewidth]{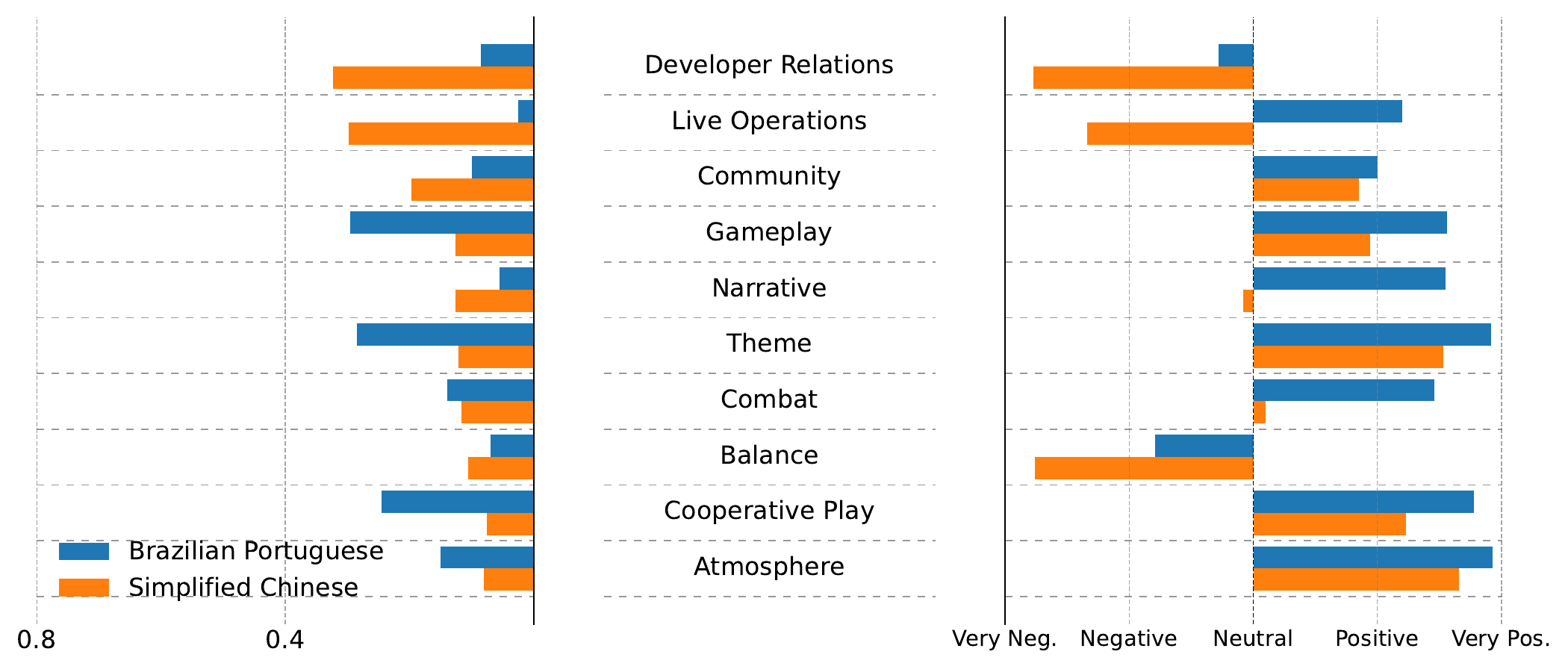}
  \captionof{figure}{HELLDIVERS 2}
  \label{fig:Helldivers2}
\end{minipage}
\end{center}

\begin{center}
\begin{minipage}{\linewidth}
  \centering
  \includegraphics[width=0.8\linewidth]{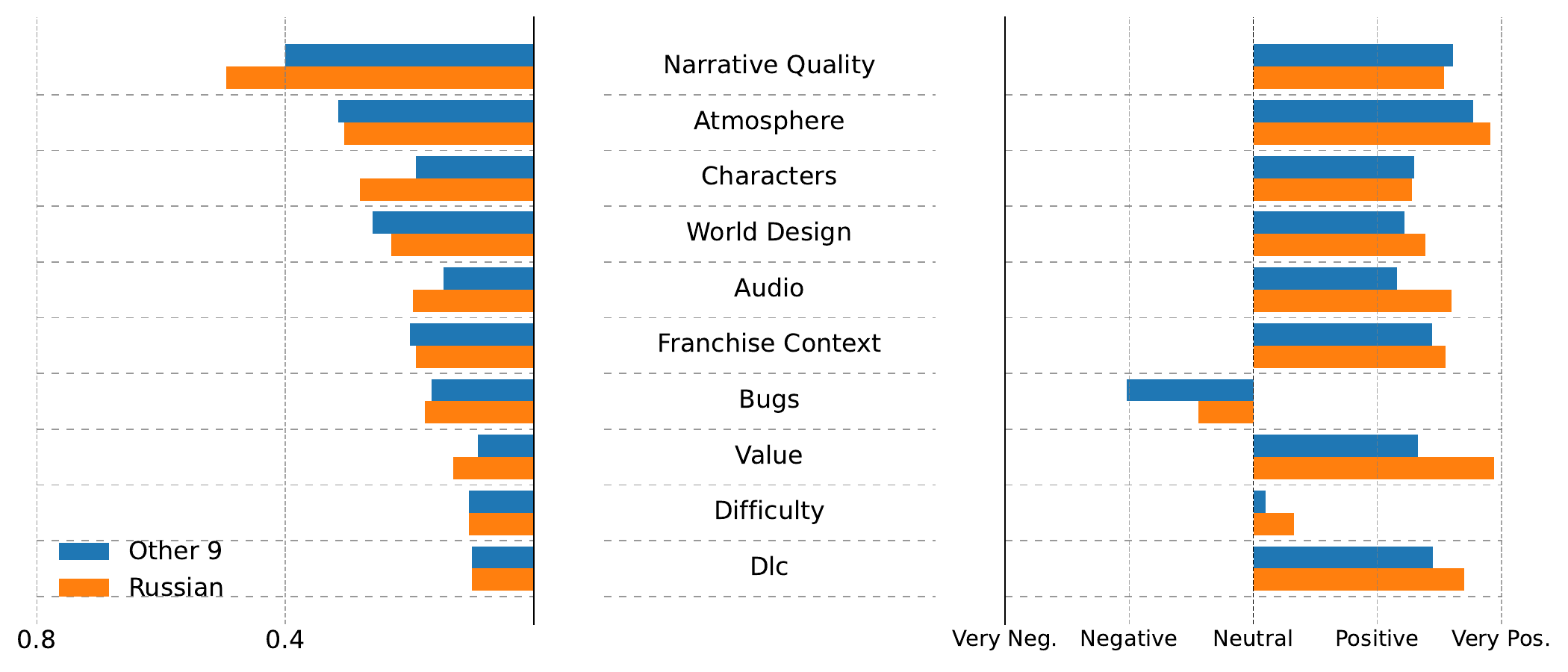}
  \captionof{figure}{Metro Exodus}
  \label{fig:MetroExodus}
\end{minipage}
\end{center}

\begin{center}
\begin{minipage}{\linewidth}
  \centering
  \includegraphics[width=0.8\linewidth]{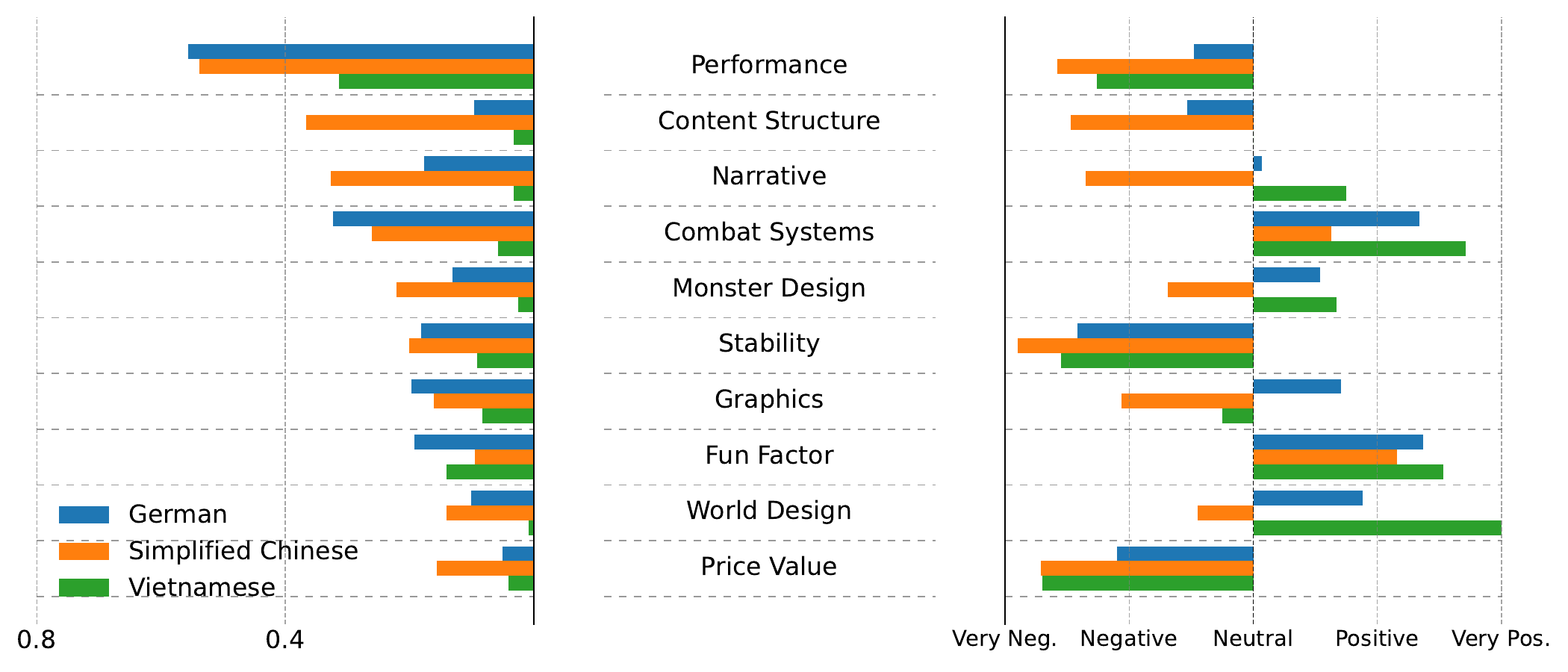}
  \captionof{figure}{Monster Hunter Wilds}
  \label{fig:MonsterHunterWilds}
\end{minipage}
\end{center}

\begin{center}
\begin{minipage}{\linewidth}
  \centering
  \includegraphics[width=0.8\linewidth]{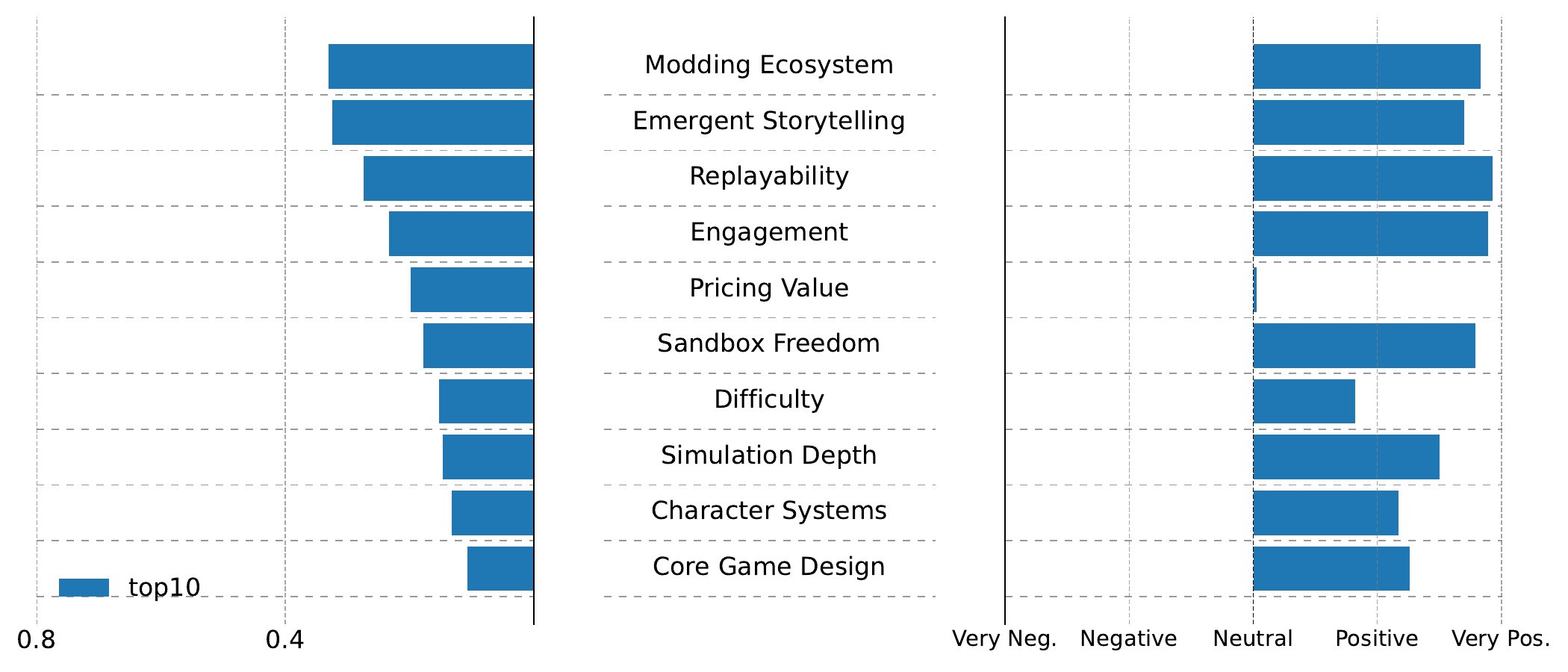}
  \captionof{figure}{RimWorld}
  \label{fig:rimworld}
\end{minipage}
\end{center}

\begin{center}
\begin{minipage}{\linewidth}
  \centering
  \includegraphics[width=0.8\linewidth]{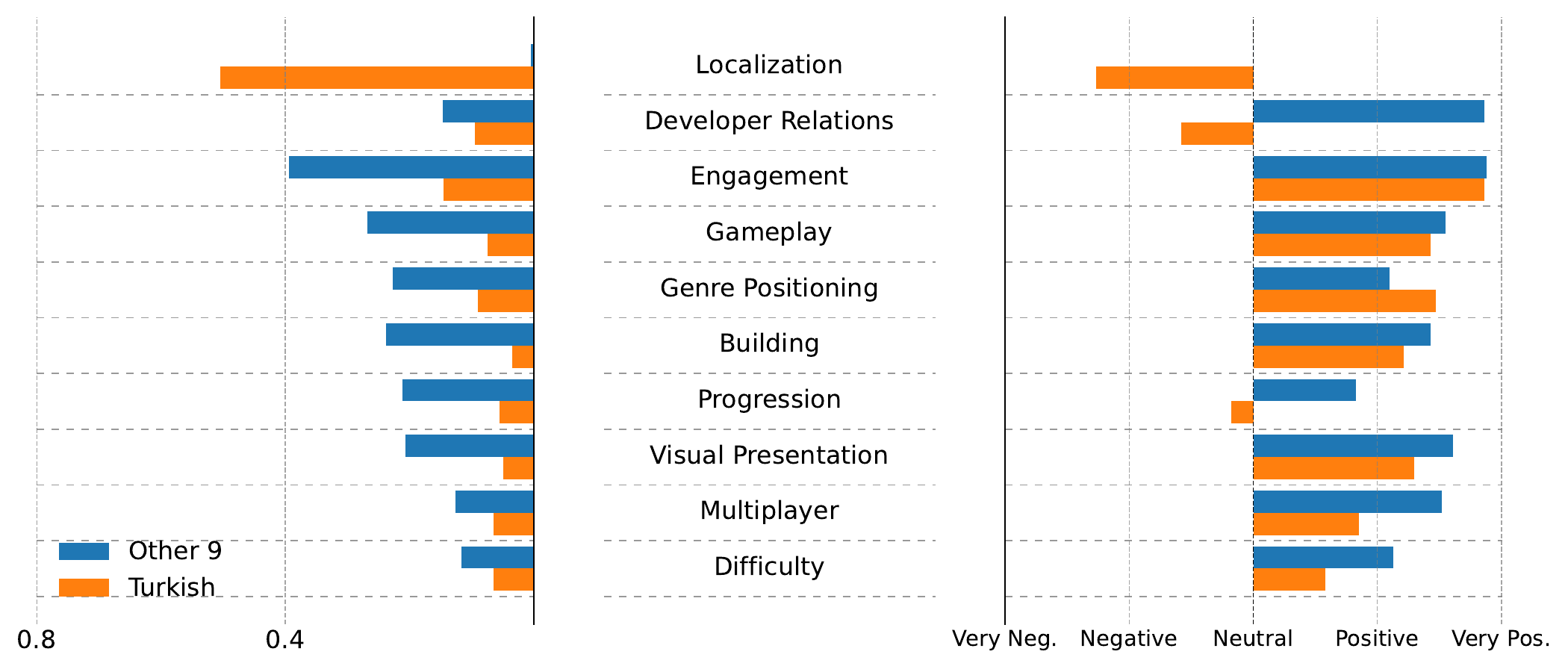}
  \captionof{figure}{Satisfactory}
  \label{fig:Satisfactory}
\end{minipage}
\end{center}

\begin{center}
\begin{minipage}{\linewidth}
  \centering
  \includegraphics[width=0.8\linewidth]{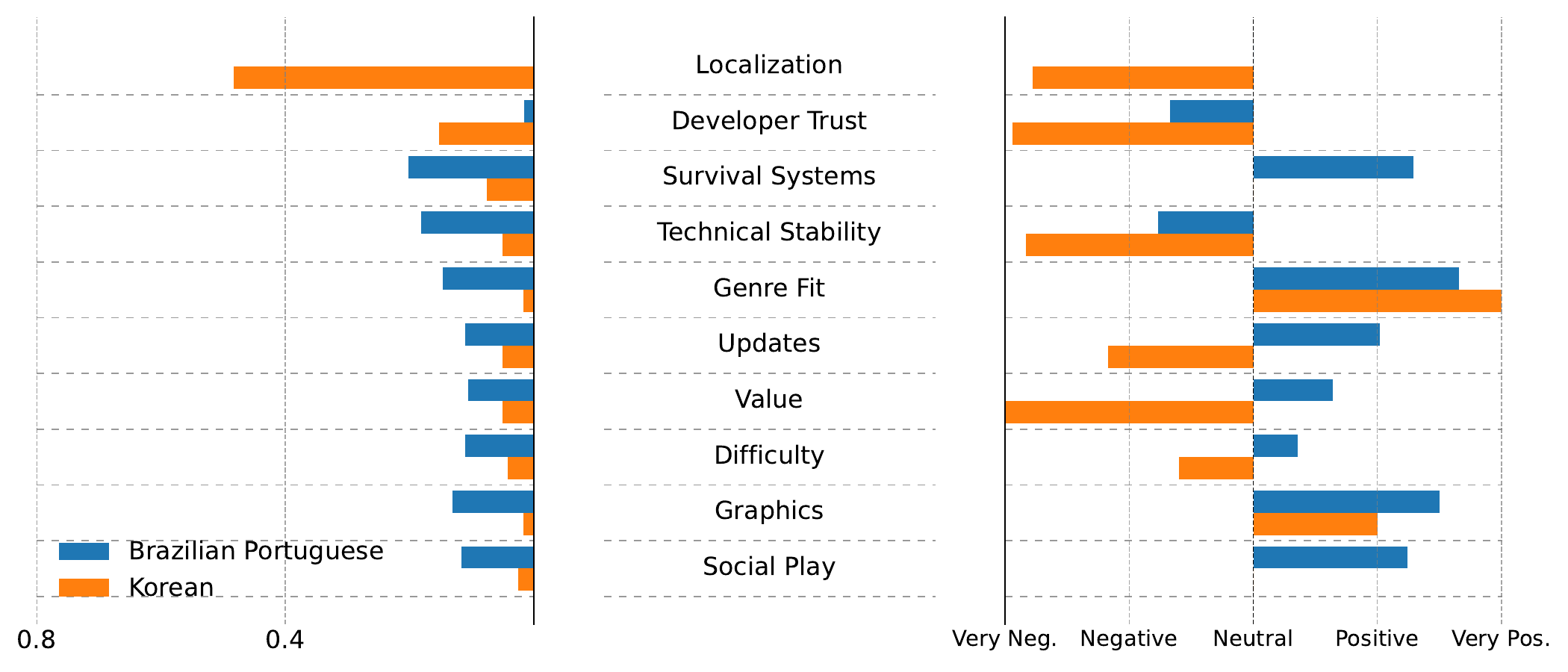}
  \captionof{figure}{SCUM}
  \label{fig:SCUM}
\end{minipage}
\end{center}

\begin{center}
\begin{minipage}{\linewidth}
  \centering
  \includegraphics[width=0.8\linewidth]{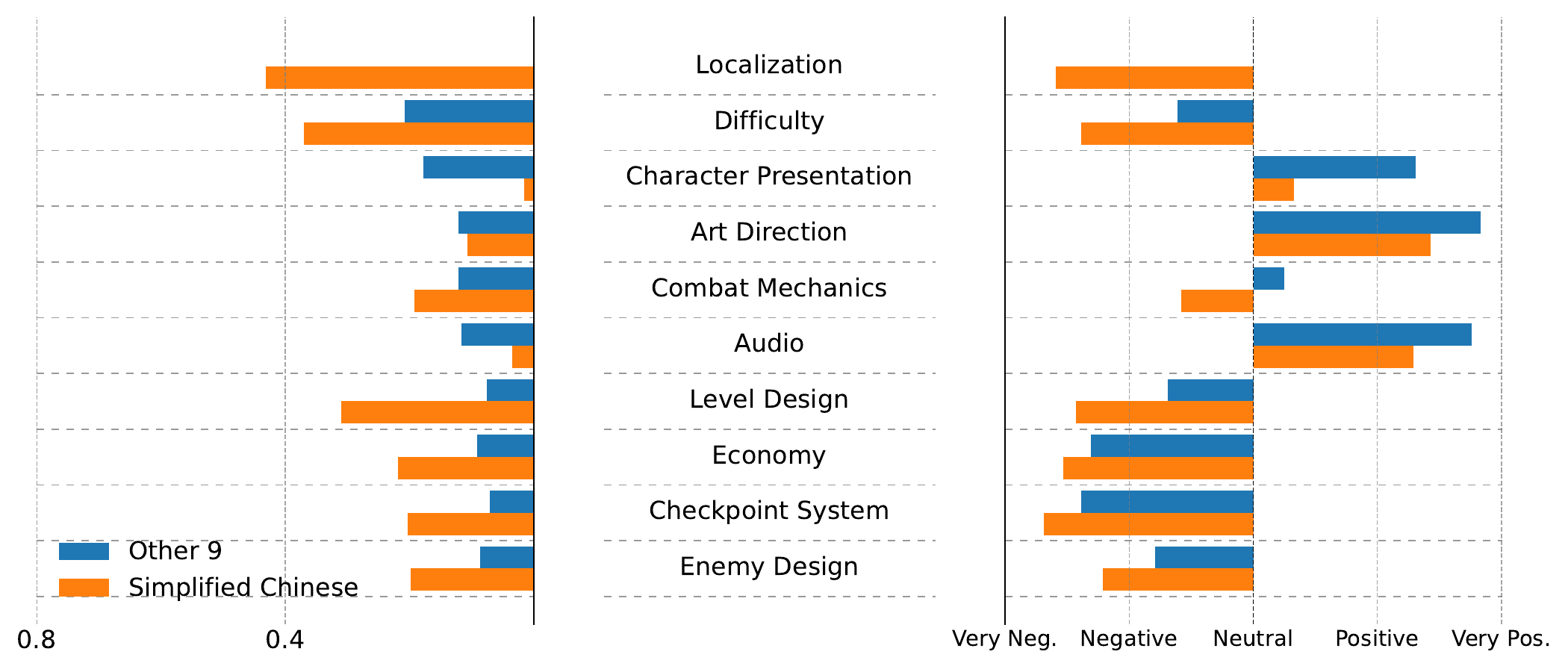}
  \captionof{figure}{Hollow Knight: Silksong}
  \label{fig:silksong}
\end{minipage}
\end{center}

\begin{center}
\begin{minipage}{\linewidth}
  \centering
  \includegraphics[width=0.8\linewidth]{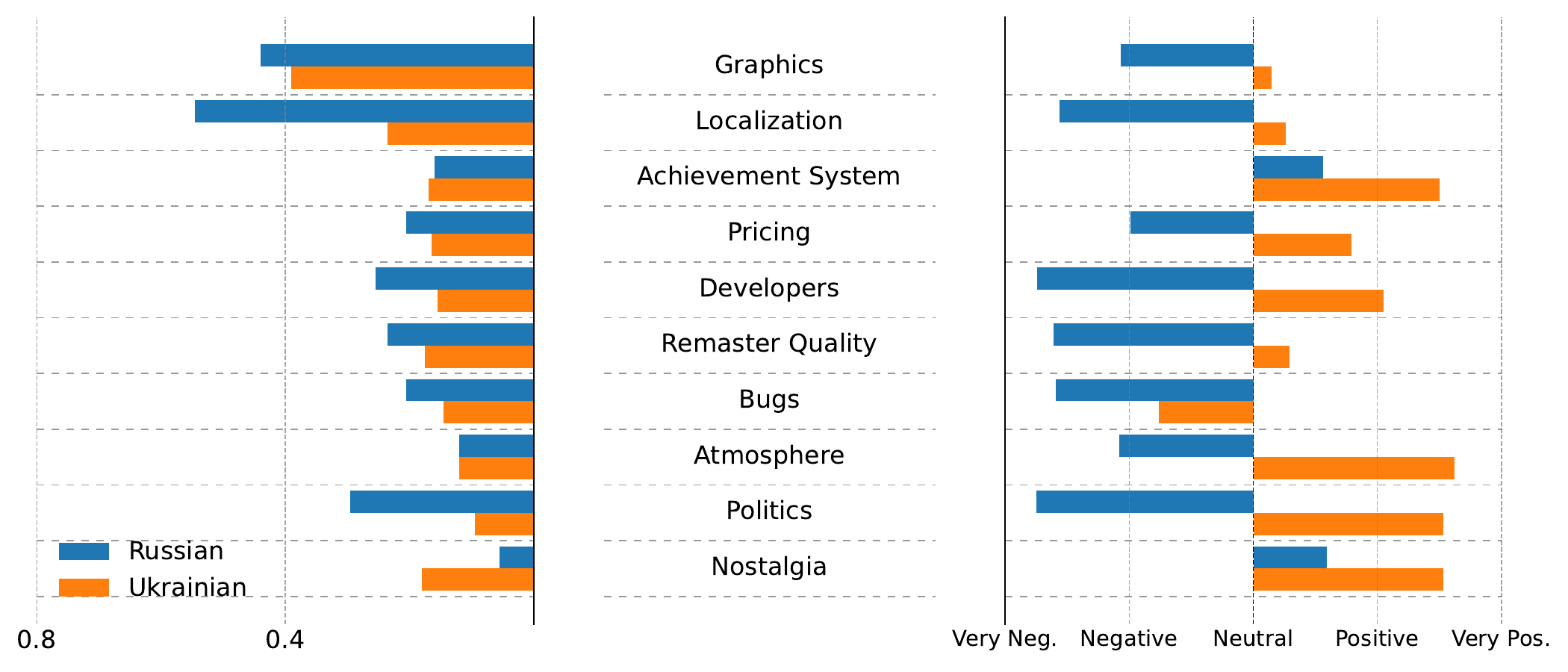}
  \captionof{figure}{S.T.A.L.K.E.R.: Shadow of Chornobyl}
  \label{fig:STALKER}
\end{minipage}
\end{center}

\begin{center}
\begin{minipage}{\linewidth}
  \centering
  \includegraphics[width=0.8\linewidth]{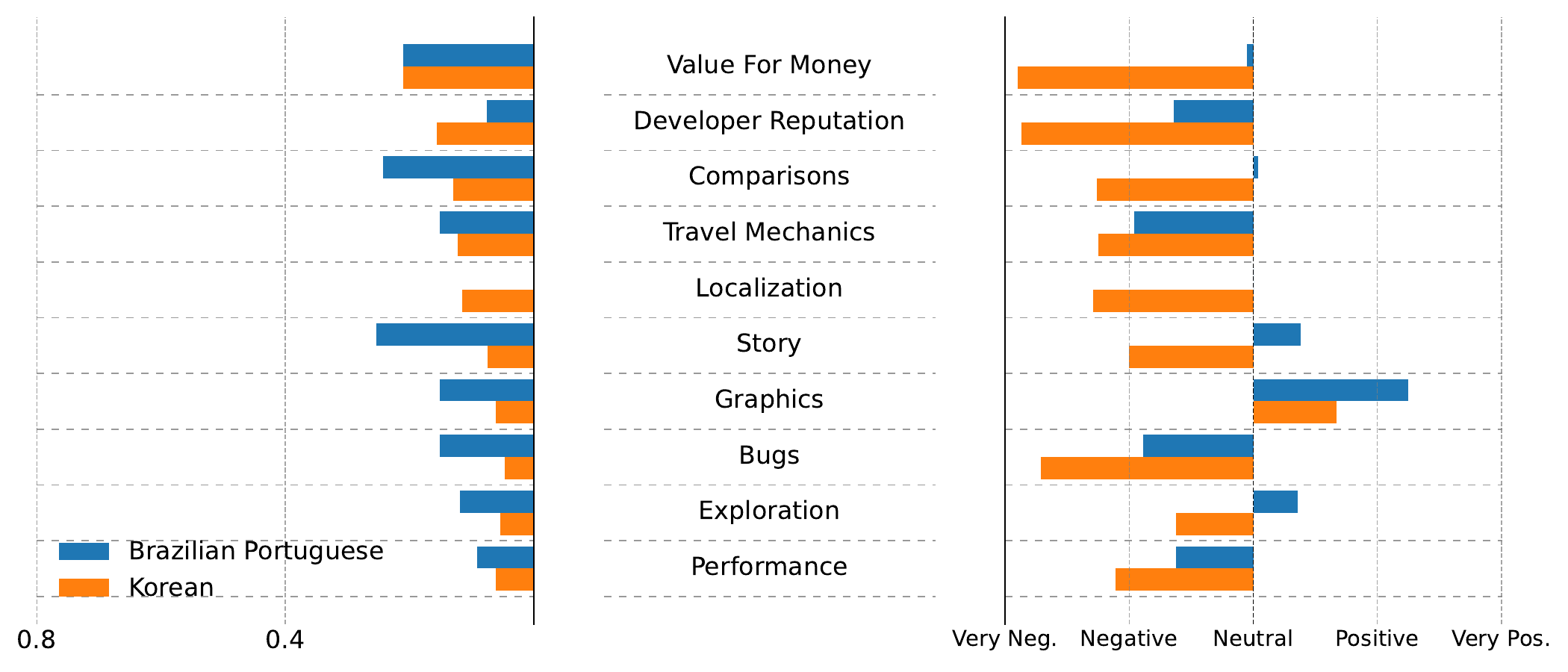}
  \captionof{figure}{Starfield}
  \label{fig:Starfield}
\end{minipage}
\end{center}

\begin{center}
\begin{minipage}{\linewidth}
  \centering
  \includegraphics[width=0.8\linewidth]{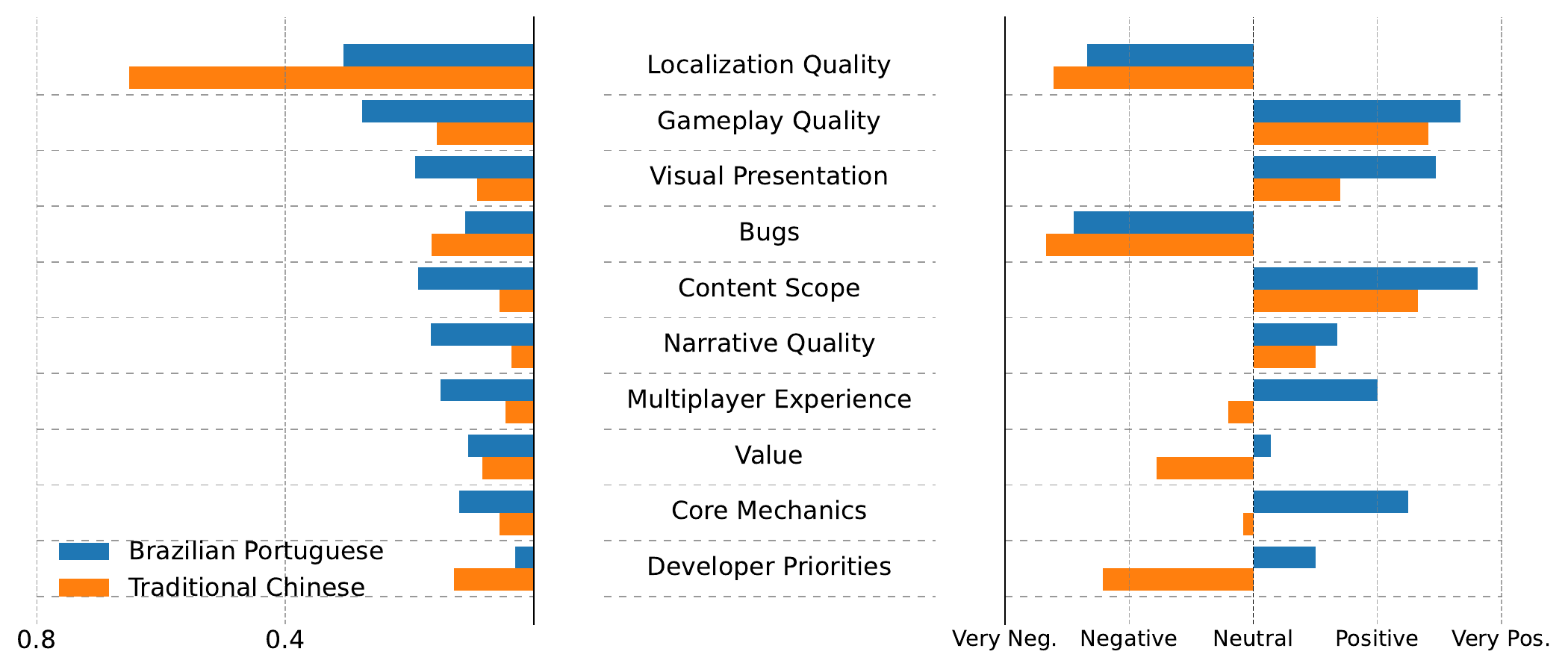}
  \captionof{figure}{Sun Haven}
  \label{fig:SunHaven}
\end{minipage}
\end{center}

\begin{center}
\begin{minipage}{\linewidth}
  \centering
  \includegraphics[width=0.8\linewidth]{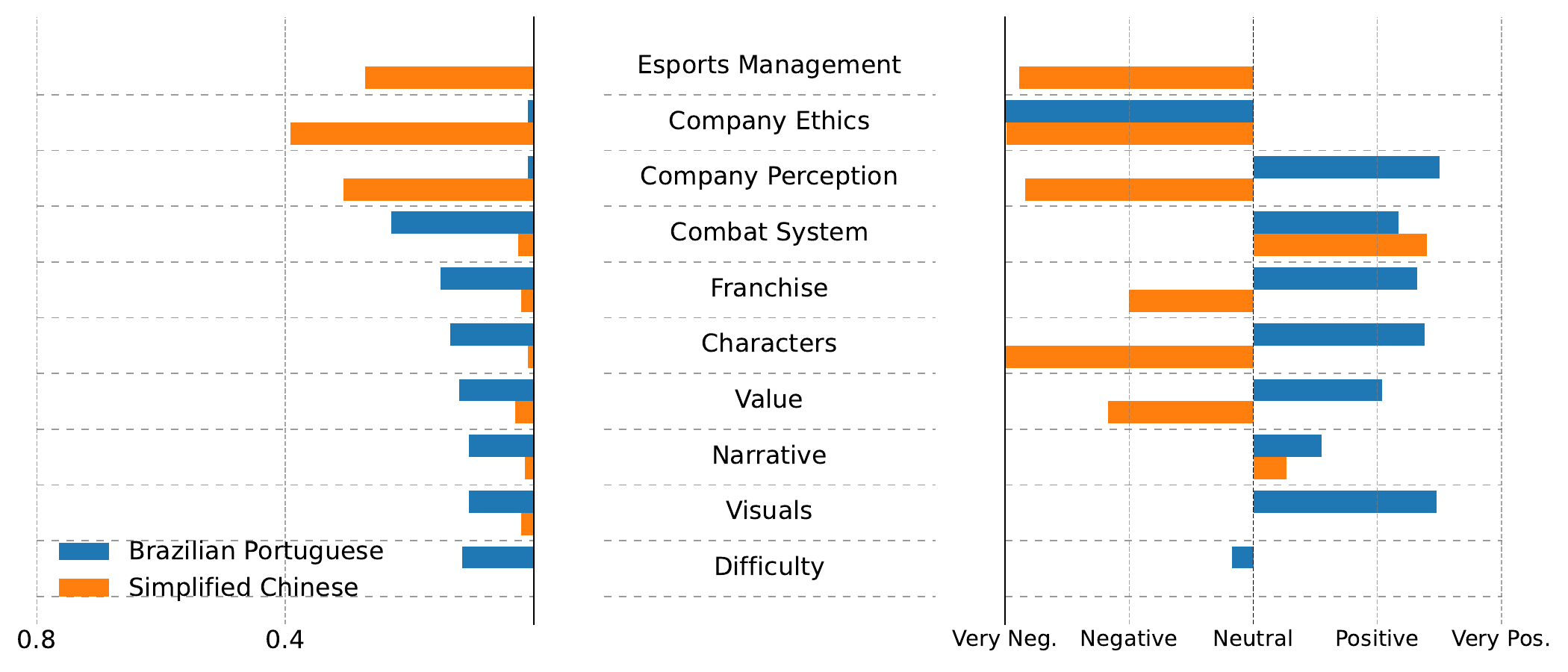}
  \captionof{figure}{TEKKEN 7}
  \label{fig:TEKKEN7}
\end{minipage}
\end{center}

\begin{center}
\begin{minipage}{\linewidth}
  \centering
  \includegraphics[width=0.8\linewidth]{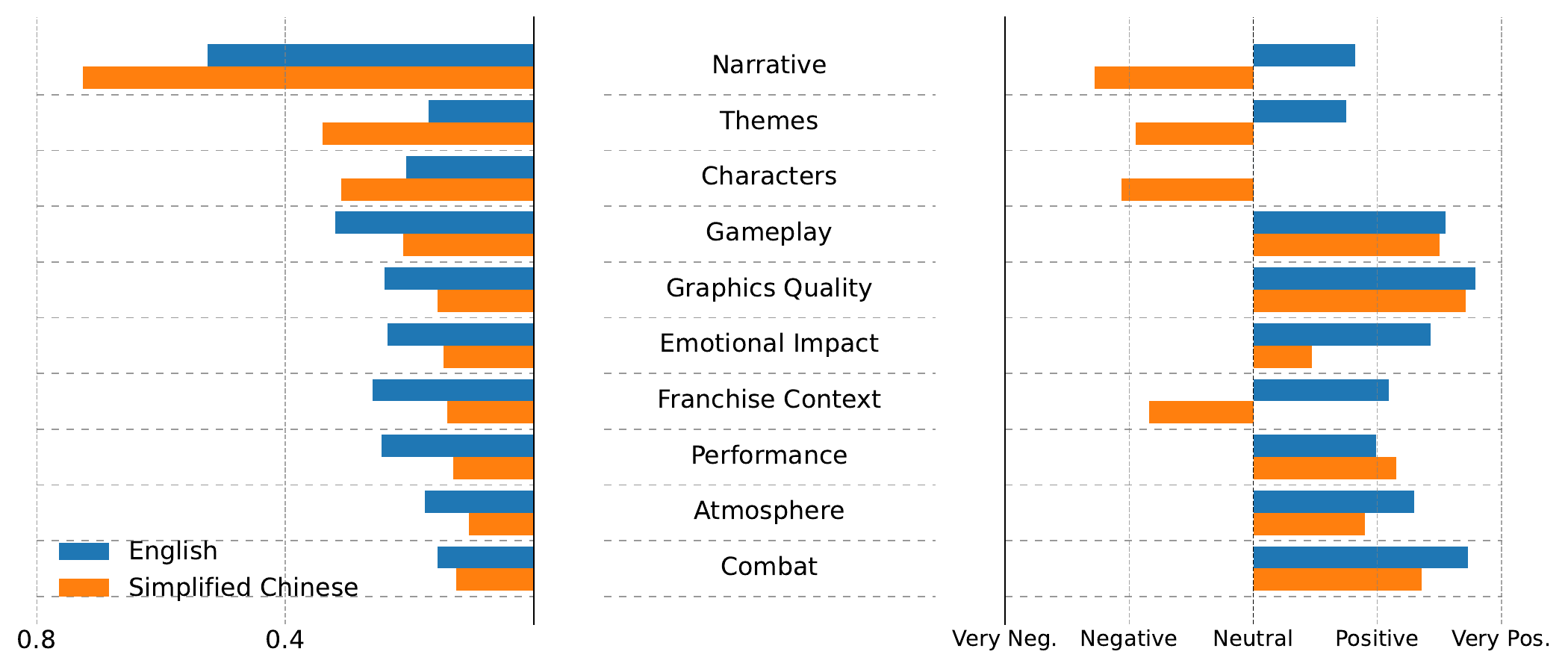}
  \captionof{figure}{The Last of Us Part II Remastered}
  \label{fig:thelastofus}
\end{minipage}
\end{center}

\begin{center}
\begin{minipage}{\linewidth}
  \centering
  \includegraphics[width=0.8\linewidth]{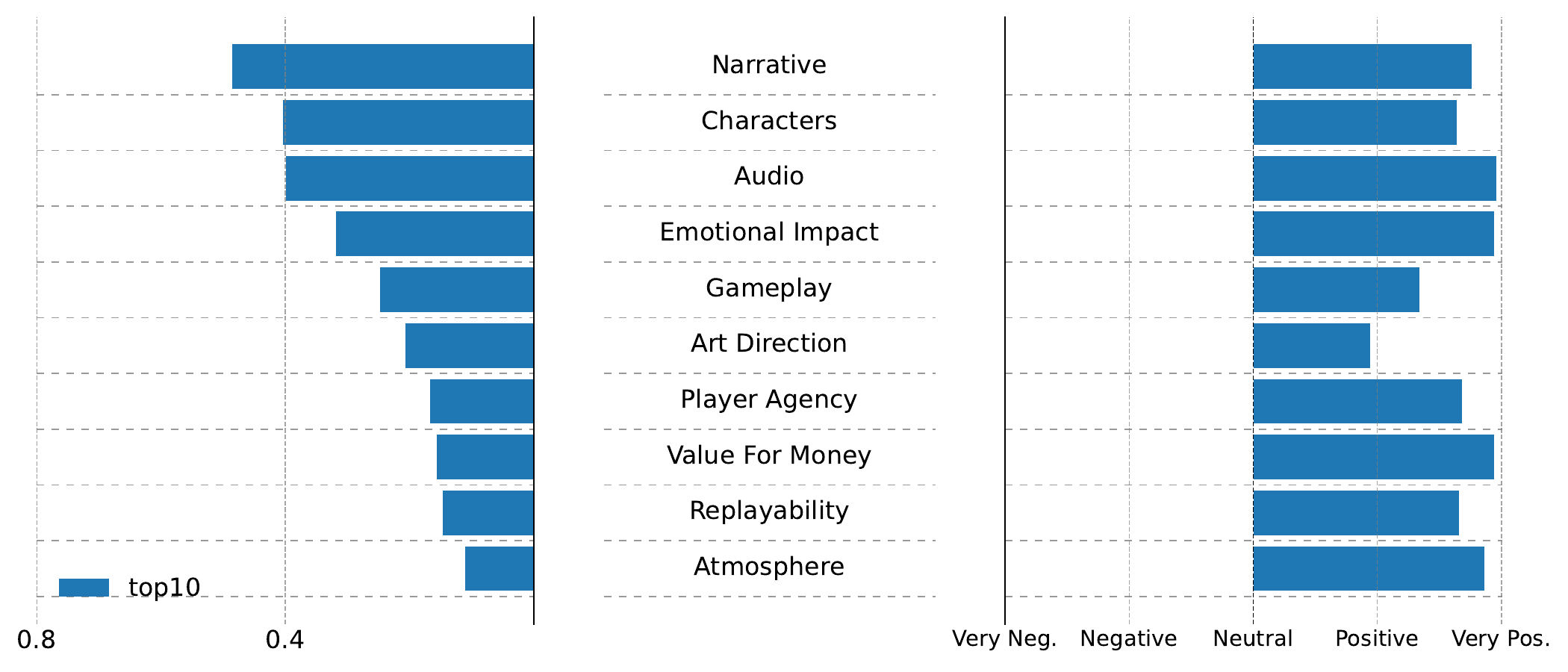}
  \captionof{figure}{Undertale}
  \label{fig:Undertale}
\end{minipage}
\end{center}